\documentclass[11pt]{article}

\usepackage{cite}
 \usepackage{subfigure}
\usepackage{multirow}
\usepackage{helvet}
\usepackage{amsmath}
\usepackage{amssymb}
\usepackage{setspace}
\usepackage{graphicx}
\usepackage{empheq}
\usepackage{soul}
\usepackage{tabularx}
\usepackage{array}
\usepackage{float}
\usepackage{braket}
\usepackage[utf8]{inputenc}
\usepackage{url}
\usepackage{hyperref}
\usepackage{slashed}
\usepackage[font=footnotesize,labelfont=bf]{caption}
\usepackage{color}

\def\be{\begin{equation}}
\def\ee{\end{equation}}

\usepackage[a4paper,top=3cm,bottom=3cm,left=2.5cm,right=2.5cm,bindingoffset=0mm]{geometry}

\begin{document}

\begin{center}
{\Large \bf Exciting Ions: a Systematic Treatment of Ultraperipheral\\ \vspace*{0.2cm} Heavy Ion Collisions with Nuclear Breakup}

\vspace*{1cm}

{\sc 
L.A.~Harland-Lang%
\footnote[1]{email: l.harland-lang@ucl.ac.uk}
}
\vspace*{0.5cm}

{\small\sl
  Department of Physics and Astronomy, University College London, London, WC1E 6BT, UK

}

\begin{abstract}
\noindent We present an updated theoretical treatment of ultraperipheral collisions (UPCs) of heavy ions, within the \texttt{SuperChic} Monte Carlo generator. This in particular accounts for mutual ion excitation through additional photon exchanges between the colliding ions. This effect occurs frequently in UPCs, and indeed can be (and has been) measured in data through the use of zero degree calorimeter (ZDC) detectors installed in the far forward region. The theoretical approach presented here accounts for the non--trivial and non--negligible impact such ion dissociation has on the measured cross sections and  distributions of the produced particles in the central detectors. This builds on previous work, whereby the survival factor probability of no additional inelastic ion--ion scattering due to the strong interaction, and its kinematic impact, are also accounted for within the same overall framework. We compare to data from ATLAS and CMS at the LHC, and STAR at RHIC, and find in general encouraging agreement for a range of observables and ZDC neutron tags, with some room for further improvement, suggesting the inclusion of higher order QED effects and/or tuning of the the $\gamma A \to A^*$ cross section may be desirable. Overall, this gives confidence in the approach considered here and for applications to new phenomena within and beyond the Standard Model.
\end{abstract}

\end{center}

\section{Introduction}

The high energy collision of heavy ions is well established as a testing ground for the strong interaction in extreme regimes, see e.g.~\cite{Braun-Munzinger:2015hba,Busza:2018rrf,ALICE:2022wpn} for  reviews. However the ions themselves carry significant electric charges $Z$ and hence can act as intense sources of electromagnetic radiation, or in the language of particle physics, photon--photon collisions. This is in particular the case for so--called ultraperipheral collisions (UPCs), where the impact parameter separation of the ions is significantly larger than the range of QCD, with the ions remaining intact after the collision. In such a case an object of interest can be produced by a purely QED interaction, with no additional particle activity present in the detector.

This production mechanism is a key ingredient in the LHC precision  and discovery programme, providing a unique probe of physics within and beyond the SM, see e.g.~\cite{Bruce:2018yzs} for a review.  One topical example is the case of light--by--light (LbyL) scattering, $\gamma\gamma \to \gamma\gamma$, for which the first ever direct evidence was presented by ATLAS in the UPC channel~\cite{ATLAS:2017fur} and subsequently by CMS~\cite{CMS:2018erd} (see~\cite{ATLAS:2019azn} for the first observation and~\cite{ATLAS:2020hii} for further analyses). This is in general sensitive to a range of new physics phenomena,  with the possible $s$--channel contribution from the decay of an axion--like--particle (ALP) being particularly relevant to UPCs; indeed, the tightest available bounds on such states in the 10--100 GeV mass region have been placed by both the ATLAS~\cite{ATLAS:2020hii} and CMS~\cite{CMS:2018erd} collaborations (see~\cite{Harland-Lang:2022end} for a recent study in $pp$ interactions). Even within the SM, possible sensitivity to tetraquark states has been discussed in~\cite{Biloshytskyi:2022dmo}. A further topical case is the potential to use the UPC channel as a probe of the anomalous magnetic moment of the $\tau$ lepton via $\gamma\gamma \to \tau \tau$ process~\cite{Beresford:2019gww,Dyndal:2020yen}. Recent observations of this process by both ATLAS~\cite{ATLAS:2022ryk} and CMS~\cite{CMS:2022arf} have already placed competitive constraints on this, with further data to come.  

The theoretical framework required to model such UPC processes has been the focus of much study (see e.g.~\cite{Bertulani:1987tz} for an early and~\cite{Klein:2020fmr} for a recent review) and a range of Monte--Carlo event generators are available~\cite{Klein:2016yzr,Harland-Lang:2018iur,Burmasov:2021phy,Shao:2022cly}. However, while the basic ingredients are universal to these,  the specific implementation (as well processes considered) is not. A particularly important element is the `survival factor' probability that the colliding ions do not interact strongly, leading to a high multiplicity event more typical of standard central ion--ion collisions, where the QED production mechanism is not so straightforwardly isolated. Now, provided the initiating photons are coherently emitted from the ions, their typical $Q^2$ is very low, which precisely corresponds to the larger impact parameter separations where  QCD interactions are suppressed. However, they are not negligible, and a precise account of this is mandatory. The survival factor in particular depends on the specific process under consideration as well as the kinematics of the produced object. The first and so far only complete treatment of this for the case of UPCs was presented in~\cite{Harland-Lang:2018iur} (see also~\cite{Harland-Lang:2021ysd}) and is implemented in the \texttt{SuperChic} Monte Carlo (MC) generator~\cite{SuperCHIC}.

A complete treatment of the survival factor, as well as all other  elements of the theoretical calculation, is crucial to taking advantage of this UPC process, where a key motivation depends on a precise understanding of the underlying photon--initiated production process. A useful testing ground for this is the production of electron and muon pairs in UPCs, which has been measured  at the LHC~\cite{ALICE:2013wjo,ATLAS:2020epq,ATLAS:2022srr,CMS:2020skx} and RHIC~\cite{STAR:2019wlg}. The ATLAS data~\cite{ATLAS:2020epq,ATLAS:2022srr} are in particular compared to the results of \texttt{SuperChic}, with generally good agreement seen, but with some systematic differences observed. This is discussed in detail in~\cite{Harland-Lang:2021ysd}, where the potential importance of higher order QED corrections is emphasised.

An important process of this type is due to additional photon exchanges between the colliding ions, whereby one or both ions can be excited into a higher energy state that subsequently decays by emitting a single or multiple neutrons. The dominant such excited state is known as giant dipole resonance (GDR), although higher excitations are also present, see e.g.~\cite{Berman:1975tt,Broz:2019kpl,Bertulani:2005ru,Baltz:2007kq,Klein:2020fmr,Baltz:2009jk}. Thus, in the presence of such exchanges the ions no longer remain intact, although as there is no colour exchange between the ions and the impact parameter can remain large, the event signature in the central detector remains the same. Moreover, these exchanges can be assumed to factorize from the underlying $\gamma\gamma \to X$ process, and hence the  cross section and distributions of the UPC process are insensitive to this, provided these are inclusive with respect to such ion dissociation. Nonetheless, this is not always true, and one can in particular measure these ion dissociation processes via zero degree calorimeter (ZDC) detectors, which have been used in UPC measurements at ATLAS~\cite{ATLAS:2020epq,ATLAS:2022srr}, CMS~\cite{CMS:2020skx} and STAR~\cite{STAR:2019wlg}. These are designed to detect neutral particles produced at very forward rapidities and so can tag events with or without additional neutron production. As ion excitation typically results in the emission of relatively low momenta neutrons in the ion rest frame, these are heavily boosted in the lab frame, and hence such ion dissociation can be measured by tagging these neutrons in the ZDCs. In other words, this is an observable effect that should be modelled appropriately.

An important element of modelling this process is, as with the case of the survival factor, to fully account for the appropriate process and kinematic dependence. The treatment of the latter effect has in particular been found to be incomplete in the  \texttt{Starlight} MC generator~\cite{Baltz:2009jk} implementation by the CMS data~\cite{CMS:2020skx} on the impact of ion dissociation on the acoplanarity of the dimuon system in UPCs. Theoretical calculations that do account for this effect~\cite{Brandenburg:2020ozx} on the other hand describe the trend of the CMS data (see e.g.~\cite{Li:2019yzy,Klein:2020jom,Wang:2021kxm,Mazurek:2021ahz,Wang:2022gkd} for other studies), but without providing a full MC treatment. In this paper, we therefore extend the \texttt{SuperChic} MC generator to account for mutual ion dissociation, including its full process and kinematic dependence, as well as the interplay with the survival factor probability. We compare to data from ATLAS, CMS and STAR, and find overall rather good agreement, as well as general consistency with the theoretical results of e.g.~\cite{Brandenburg:2020ozx}. Therefore, this extension will allow predictions for UPC production to be compared more directly with what is measured experimentally, given as we will see the ion dissociation probability is certainly non--negligible and does impact on the corresponding kinematic distributions in the presence of ZDC tags.

The outline of this paper is as follows. In Section~\ref{sec:key} we present the key ingredients in the model we apply for UPCs. In Section~\ref{sec:surv} we discuss the ion--ion survival factor, and how this is accounted for. In Section~\ref{sec:iondiss} we describe how this framework can account for the effect of mutual ion dissociation. In Section~\ref{sec:results} we compare to a range of data from ATLAS, CMS and STAR. Finally, in Section~\ref{sec:conc} we conclude.

\section{Theory}\label{sec:theory}

\subsection{Key ingredients}\label{sec:key}

The basic formalism follows that described in for example~\cite{Harland-Lang:2018iur}. Omitting the survival factor for now, the cross section can be written as 
\be\label{eq:csn}
\sigma =\frac{1}{2s}\int {\rm d} x_1 {\rm d}x_2 {\rm d}^2 q_{1\perp}{\rm d}^2 q_{2\perp} {\rm d}\Gamma\frac{1}{\tilde{\beta}} |T(q_{1\perp},q_{2\perp}) |^2\delta^{4}(q_1+q_2-k)\;,
\ee
  where $x_i$ and $q_{i\perp}$ are the photon momentum fractions (see~\cite{Harland-Lang:2019zur} for precise definitions) with respect to the parent ion beams and the photon transverse momenta, respectively.  The photons have momenta $q_{1,2}$, with $q_{1,2}^2 = -Q_{1,2}^2$, and we consider the production of a system of 4--momentum $k = q_1 + q_2 = \sum_{j=1}^N k_j$ of $N$ particles, where ${\rm d}\Gamma = \prod_{j=1}^N {\rm d}^3 k_j / 2 E_j (2\pi)^3$ is the standard phase space volume.  $\tilde{\beta}$ is as defined in~\cite{Harland-Lang:2019zur} and $s$ is the ion--ion squared c.m.s. energy.
  
In \eqref{eq:csn}, $T$ is the process amplitude, and is given by
\be\label{eq:tq1q2}
T(q_{1\perp},q_{2\perp}) = \mathcal{N}_1 \mathcal{N}_2 \,q_{1\perp}^\mu q_{2\perp}^\nu V_{\mu\nu}\;,
\ee  
where $V_{\mu\nu}$ is the $\gamma^*\gamma^* \to X$ vertex, i.e. the amplitude that in the on--shell case would couple to the photon polarization vectors $\epsilon$. The normalization factors are 
\be\label{eq:ni}
\mathcal{N}_i = \frac{2\alpha(Q_i^2)^{1/2}}{ x_i}\frac{F_p(Q_i^2)G_E(Q_i^2)}{Q_i^2}\;.
\ee
where $F_p^2(Q^2)$ is the squared form factor of the ion and 
\be\label{eq:qi}
Q^2_i=\frac{q_{i_\perp}^2+x_i^2 m_{A_i}^2}{1-x_i}\;.
\ee
The squared form factor is given in terms of the proton density in the ion, $\rho_p(r)$, which is well described by the Woods--Saxon distribution~\cite{Woods:1954zz}
\be\label{eq:rhop}
\rho_p(r)= \frac{\rho_0}{1+\exp{\left[(r-R)/d\right]}}\;,
\ee
where the skin thickness $d \sim 0.5-0.6$ fm, depending on the ion, and the radius $R \sim A^{1/3}$. To be precise, for Pb ions we take the experimentally determined values~\cite{Tarbert:2013jze}
\begin{align}\nonumber
R_p &= 6.680\, {\rm fm}\;, &d_p &= 0.447 \, {\rm fm}\;,\\ \label{eq:pbpar}
R_n &= (6.67\pm 0.03)\, {\rm fm}\;, &d_n &= (0.55 \pm 0.01) \, {\rm fm}\;.
\end{align}
For Au ions we scale these values of $R_{n,p}$ by $A^{1/3}$, while keeping the $d_{p,n}$ fixed. The density $\rho_0$ is set by requiring that
\be
\int {\rm d}^3  r\,\rho_p(r) = Z\;.
\ee
The ion form factor is then simply given by the Fourier transform
\be\label{eq:Fp}
F_p(|{\mathbf q}|) = \int {\rm d}^3r \, e^{i {\mathbf q}\cdot {\mathbf r}}  \rho_p(r)\;,
\ee
in the rest frame of the ion; in this case we have ${\mathbf q}^2 = Q^2$, so that written covariantly this corresponds to the $F_p(Q^2)$ which appears in \eqref{eq:ni}.

In \eqref{eq:ni} $G_E$ corresponds to the form factor of the protons within the ion; numerically this has a negligible impact, as the ion form factor falls much more steeply, however we include this for completeness.   
To be precise, $G_E$ is proton EM `Sachs' form factor, given by
\begin{equation}\label{eq:dip}
G_E^2(Q_i^2)=\frac{1}{\left(1+Q^2_i/0.71 {\rm GeV}^2\right)^4}\;,
\end{equation}
in the dipole approximation. In this work we do not use the dipole approximation but rather, as in~\cite{Harland-Lang:2020veo}, the fit from the A1 collaboration~\cite{Bernauer:2013tpr}. 
 
We note that in \eqref{eq:tq1q2} we can see that the amplitude depends in general on a range of other kinematic variables (the photon momentum fractions $x_i$, the kinematics of the produced final state and so on). These are always implied, but omitted in the arguments of $T$ for brevity. We keep the dependence on the photon transverse momentum $q_{i_\perp}$ explicit for clarity, in particular when discussing the survival factor in the following section.
 
In the high energy limit, and neglecting the off--shellness of the initial--state photons in the $\gamma\gamma \to X$ process, the above expression reduces to a well known result of the equivalent photon approximation~\cite{Budnev:1974de}, which is formulated purely at the cross section level. To be precise, we can rewrite \eqref{eq:csn} as 
\be\label{eq:csepa}
{\rm d} \sigma = \int \frac{ {\rm d} x_1 }{x_1} \frac{ {\rm d} x_2 }{x_2}n(x_1) n(x_2) {\rm d}\sigma_{\gamma\gamma \to X}\;,
\ee
in these limits. The flux $n(x_i)$ is given by
\be
n(x_i) =  \int {\rm d}^2 q_{i_\perp}  | {\mathbf N}(x_i,q_{i_\perp})|^2 \;,
\ee
where
\be\label{eq:nibf}
{\mathbf N}(x_i,q_{i_\perp}) =  \frac{ \alpha(Q_i^2)^{1/2}}{\pi} \frac{{\mathbf q}_{i_\perp}}{q_{i_\perp}^2 + x_i^2 m_A^2}F_p(Q_i^2)G_E(Q_i^2)\;.
\ee
Equivalently the above formulation follows simply from the structure function expression in e.g.~\cite{Harland-Lang:2019eai}, once we drop the subleading magnetic form factor of the ion, which is suppressed by a factor of $\sim Z$, and take the high energy limit. However, formulating things as in \eqref{eq:csn} allows us to fully keep track of the non--zero photon off--shellness, while working in terms of the amplitude $T$ allows us to include the survival factor, and ion dissociation effects, in an appropriate way. We consider this in the next section. 
  
Finally, we note that, in the EPA limit it was shown in~\cite{Harland-Lang:2010ajr} that \eqref{eq:tq1q2} can be recast via
\begin{align}
q_{1_\perp}^i q_{2_\perp}^j V_{ij} =\begin{cases} &-\frac{1}{2} ({\bf q}_{1_\perp}\cdot {\bf q}_{2_\perp})(\mathcal{M}_{++}+\mathcal{M}_{--})\;\;(J^P_z=0^+)\\ 
&-\frac{i}{2} |({\bf q}_{1_\perp}\times {\bf q}_{2_\perp})|(\mathcal{M}_{++}-\mathcal{M}_{--})\;\;(J^P_z=0^-)\\ 
&+\frac{1}{2}((q_{1_\perp}^x q_{2_\perp}^x-q_{1_\perp}^y q_{2_\perp}^y)+i(q_{1_\perp}^x q_{2_\perp}^y+q_{1_\perp}^y q_{2_\perp}^x))\mathcal{M}_{-+}\;\;(J^P_z=+2^+)\\ 
&+\frac{1}{2}((q_{1_\perp}^x q_{2_\perp}^x-q_{1_\perp}^y q_{2_\perp}^y)-i(q_{1_\perp}^x q_{2_\perp}^y+q_{1_\perp}^y q_{2_\perp}^x))\mathcal{M}_{+-}\;\;(J^P_z=-2^+)
\end{cases}\label{eq:Agen}
\end{align}
where $\mathcal{M}_{\pm \pm}$ corresponds to the $\gamma(\pm) \gamma(\pm) \to X$ helicity amplitudes and $J^P_z$ the spin--parity of the corresponding $\gamma\gamma$ configuration. While we do not explicitly make use of this formula in our calculation, it will be useful to consider the above expression later on when discussing the comparison to the STAR data~\cite{STAR:2019wlg}.
  
 \subsection{The Survival Factor}\label{sec:surv}
 
The inclusion of the survival factor closely follows the description in~\cite{Harland-Lang:2018iur}, although here we present this in a somewhat different way, in order to facilitate the discussion when we consider mutual ion dissociation. In the high energy limit, and neglecting the off--shellness of the initial--state photons in the $\gamma\gamma \to X$ process kinematics we can write
\begin{align}\nonumber
\sigma &=\frac{1}{2s}\int {\rm d} x_1 {\rm d}x_2 {\rm d}^2 q_{1\perp}{\rm d}^2 q_{2\perp} {\rm d}\Gamma\frac{1}{\tilde{\beta}} |T(q_{1\perp},q_{2\perp}) |^2\delta^{4}(q_1+q_2-k)\;,\\ \nonumber
&\approx \frac{1}{2s}\int {\rm d} x_1 {\rm d}x_2 {\rm d}\Gamma\frac{1}{\tilde{\beta}} \delta^{4}(q_1+q_2-k)\int {\rm d}^2 q_{1\perp}{\rm d}^2 q_{2\perp}|T(q_{1\perp},q_{2\perp}) |^2\;,\\
& \equiv  \frac{1}{2s}\int {\rm d} x_1 {\rm d}x_2 {\rm d}\Gamma\frac{1}{\tilde{\beta}} \delta^{4}(q_1+q_2-k) \,{\rm d} \sigma\;,
\end{align}
although we emphasise that in all calculations we make use of the full form as per \eqref{eq:csn}. Our discussion of the survival factor then concerns the object
\be
{\rm d} \sigma = \int {\rm d}^2 q_{1\perp}{\rm d}^2 q_{2\perp}|T(q_{1\perp},q_{2\perp}) |^2\;.
\ee
The physical interpretation is clearest when we move to impact parameter space, i.e. taking the Fourier transform of the amplitude that appears above to give
\be\label{eq:tqb}
\tilde{T}(b_{1\perp},b_{2\perp})=\frac{1}{(2\pi)^2} \int {\rm d}^2 q_{1\perp} {\rm d}^2 q_{2\perp} e^{-i {\mathbf q}_{1\perp}\cdot {\mathbf b}_{1\perp}}e^{i {\mathbf q}_{2\perp}\cdot {\mathbf b}_{2\perp}}T(q_{1\perp},q_{2\perp})\;,
\ee 
such that
\be
{\rm d} \sigma = \int {\rm d}^2 b_{1\perp}{\rm d}^2 b_{2\perp} |\tilde{T}(b_{1\perp},b_{2\perp})|^2\;.
\ee
The survival factor is then accounted for by considering 
\be\label{eq:sigbs2}
{\rm d} \sigma_{\rm S^2} = \int {\rm d}^2 b_{1\perp}{\rm d}^2 b_{2\perp} |\tilde{T}(b_{1\perp},b_{2\perp})|^2\,\Gamma_{A_1 A_2}(s, b_\perp) \;,
\ee
where 
$\Gamma_{A_1A_2}$ represents the probability that no inelastic scattering occurs at impact parameter $b_\perp=| {\mathbf b}_{1\perp}+{\mathbf b}_{2\perp}|$, and weights the cross section including the survival factor in the appropriate way. It is typically written in terms of the ion--ion opacity $\Omega_{A_1A_2}$ via
\be\label{eq:opac}
\Gamma_{A_1A_2}(s,b_\perp)  \equiv \exp(-\Omega_{A_1 A_2}(s,b_\perp))\;.
\ee
This is given in terms of the opacity due to nucleon--nucleon interactions, $\Omega_{nn}$, which is in turn given by a convolution of the nucleon--nucleon scattering amplitude $A_{nn}$ and the transverse nucleon densities $T_n$, see~\cite{Harland-Lang:2018iur} for a more detailed discussion. To first approximation, we have
\be\label{eq:opapprox}
e^{-\Omega_{A_1A_2}(s,b_\perp)/2} \approx \theta(b_\perp - 2 R_A)\;,
\ee
i.e. it corresponds to a requiring that the ions to do not overlap in impact parameter, although in the full calculation the opacity shows some departure from this, see~\cite{Harland-Lang:2018iur}.

At the amplitude level \eqref{eq:sigbs2} simply corresponds to replacing
\be\label{eq:tbgam}
\tilde{T}(b_{1\perp},b_{2\perp}) \to \tilde{T}(b_{1\perp},b_{2\perp})\Gamma_{A_1A_2}(s,b_\perp)^{1/2}\;,
\ee
where we will comment on the (lack of) any complex phase at the end of this section.
Moving back to transverse momentum space, we are therefore interested in
\be\label{eq:tqts2}
T_{\rm S^2}(q_{1\perp},q_{2\perp}) =\frac{1}{(2\pi)^2}\int {\rm d}^2 b_{1\perp}{\rm d}^2 b_{2\perp}\,e^{i {\mathbf q}_{1\perp}\cdot {\mathbf b}_{1\perp}}e^{-i {\mathbf q}_{2\perp}\cdot {\mathbf b}_{2\perp}}\tilde{T}(b_{1\perp},b_{2\perp})\Gamma_{A_1A_2}(s,b_\perp)^{1/2}\;,
\ee
where the `$S^2$' subscript indicates that the survival factor has now been appropriately accounted for; we then substitute this in \eqref{eq:csn} to get the cross section. 
A convenient form for the above amplitude comes from defining
\be\label{eq:pmc1}
\mathcal{P}_{A_1A_2}(s,k_\perp)\equiv \frac{1}{(2\pi)^2}\int {\rm d}^2 b_{\perp}\,e^{i {\mathbf k}_{\perp}\cdot {\mathbf b}_{\perp}} \Gamma_{A_1A_2}(s,b_\perp)^{1/2}\;,
\ee
in terms of which we have
\be
T_{\rm S^2}(q_{1\perp},q_{2\perp}) = \int  {\rm d}^2 k_{\perp}\, T(q_{1\perp}',q_{2\perp}')\,\mathcal{P}_{A_1A_2}(s,k_\perp)\;,
\ee
where $q_{1\perp}' =q_{1_\perp} - k_\perp$ and $q_{2\perp}' = q_{2\perp} + k_\perp$. We can see from \eqref{eq:opapprox} that as it stands \eqref{eq:pmc1} involves an integral that extends from $b_\perp \sim 2 R_A$ to infinity, which while formally convergent, is numerically  unstable. With this in mind we can also define
\be\label{eq:pmc2}
\mathcal{P}_{A_1A_2}'(s,k_\perp)\equiv  \frac{1}{(2\pi)^2}\int {\rm d}^2 b_{\perp}\,e^{i {\mathbf k}_{\perp}\cdot {\mathbf b}_{\perp}} \left[1-\Gamma_{A_1A_2}(s,b_\perp)^{1/2}\right]\;,
\ee
in terms of which we have
\be
T_{\rm S^2}(q_{1\perp},q_{2\perp}) =T(q_{1\perp},q_{2\perp})- \int  {\rm d}^2 k_{\perp}\, T(q_{1\perp}',q_{2\perp}')\,\mathcal{P}'_{A_1A_2}(s,k_\perp)\;.
\ee
While these two formulations are in principle completely equivalent, we can see  that \eqref{eq:pmc2} now involves an integral from the finite range $b_\perp =0$ to  $b_\perp \sim 2 R_A$, which is numerically stable.

We note that it is common to refer to the survival factor, which corresponds to the ratio of the cross section evaluated using $T_{\rm S^2}$ to that evaluated using $T$. However, it is clear from \eqref{eq:tqts2} that the ratio
\be
\frac{|T_{\rm S^2}(q_{1\perp},q_{2\perp}) |^2}{|T(q_{1\perp},q_{2\perp}) |^2}\;,
\ee
is dependent on the photon transverse momenta $q_{i_\perp}$ as well as the particular form of $T(q_{1_\perp},q_{2_\perp})$. The former dependence implies that the survival factor will modify the predicted kinematic distributions, both those directly dependent on $q_{i_\perp}$ such as acoplanarity distributions but also through \eqref{eq:qi} which couples the $q_{i_\perp}$ dependence to the photon momentum fractions $x_i$ (and hence the $\gamma\gamma$ invariant mass and rapidity). The latter dependence implies that the survival factor is dependent on the particular process under consideration. Further discussion of this can be found in e.g.~\cite{Harland-Lang:2021ysd,Harland-Lang:2020veo} and references therein. We also note that although only UPCs are considered here, one can in principle readily extend the above approach to include collisions at other centralities, that is by suitably modifying $\Gamma_{AA}$ in \eqref{eq:sigbs2} to have support for the appropriate $b_\perp$ range corresponding to the required centrality class.

Finally, one might worry in \eqref{eq:tbgam} about the uniqueness of this replacement, given one could in principle multiply by an arbitrary $b_\perp$ dependent complex phase and still give the same integrated cross section \eqref{eq:sigbs2}  evaluated in impact parameter space. This will however in general modify \eqref{eq:pmc1} and, one might worry, the corresponding predictions for the differential cross section with respect to the photon transverse momenta. The solution to this comes from noting that the individual photon transverse momenta are not observable, but only their vector sum, i.e. the transverse momentum of the measured centrally produced state. We can then write \eqref{eq:tqts2} as 
\be\label{eq:tqts2rw}
T_{\rm S^2}(q_{1\perp},q_{2\perp}) =\frac{1}{2(2\pi)^2}\int {\rm d}^2 b_{\perp}{\rm d}^2 b_{\perp}'\,e^{i {\mathbf q}_{\perp}\cdot {\mathbf b}_\perp'}e^{i {\mathbf l}_{\perp}\cdot {\mathbf b}_{\perp}}\tilde{T}(b_{1\perp},b_{2\perp})\Gamma_{A_1A_2}(s,b_\perp)^{1/2}\;,
\ee
where we have defined $\mathbf{b}_\perp' =  {\mathbf b}_{1\perp}-{\mathbf b}_{2\perp}$, with $\mathbf{b}_\perp$ defined as before, and 
\begin{align}
{\mathbf l}_\perp  &= \frac{1}{2}\left({\mathbf q}_{1\perp} - {\mathbf q}_{2\perp}\right)\;,\\
{\mathbf q}_\perp  &= \frac{1}{2}\left({\mathbf q}_{1\perp} + {\mathbf q}_{2\perp}\right)\;.
\end{align}
It  then follows straightforwardly from \eqref{eq:tqts2rw} that 
\be
\int {\rm d}^2 l_\perp |T_{\rm S^2}(q_{1\perp},q_{2\perp}) |^2 \sim \int {\rm d}^2 b_{\perp}  {\rm d}^2 \tilde{b}_{\perp}\,  \left[\Gamma_{A_1A_2}(s,b_\perp)^{1/2} \Gamma_{A_1A_2}^*(s,\tilde{b}_\perp)^{1/2}\cdots\right] \delta^{(2)}(\mathbf{b}_\perp-\tilde{\mathbf{b}}_\perp)\;,
\ee
and hence any $b_\perp$ dependent phase in \eqref{eq:tbgam} will not contribute. This can therefore  without loss of generality be simply dropped.

 \subsection{Including Mutual Ion Dissociation}\label{sec:iondiss}
 
 We include ion dissociation following the well--established formalism described in e.g.~\cite{Hencken:1995me,Baltz:1996as,Baltz:2009jk,Klein:2016yzr}. The basic idea is that the ion dissociation occurs via a single (or multiple) photon exchanges between the ions, such that coherent photon emission process on one ion side leads to ion dissociation via $\gamma A \to A^*$ on the other ion side. If the ion dissociates, it will emit some number $n$ neutrons as part of its `de--excitation'. As these are typically emitted with rather low momenta in the ion rest frame, in the lab frame these are highly boosted to forward rapidities, where they can be detected by ZDCs. The dominant nuclear excitation is a so--called `giant dipole resonance' (GDR)~\cite{Berman:1975tt,Broz:2019kpl,Bertulani:2005ru,Baltz:2007kq,Klein:2020fmr,Baltz:2009jk}, which is the lowest energy nuclear excitation and typically leads to the emission of a single neutron. However, higher excitations are possible, for which a larger number $X$ neutrons will be emitted (in addition to other particles such as pions). We therefore consider three possibilities in our calculation, namely the emission of zero, one or any number $X > 0$ neutrons; while in principle one can provide predictions for any exclusive number of neutron emissions, we will for simplicity not consider these here, beyond the single neutron case. 
 
 We also note that we will only consider mutual ion dissociation here. In general, the initial state photons in the $\gamma\gamma$ initiated process may be emitted inelastically from the ions, however this is subleading, being suppressed by a factor of $\sim Z$, so that it is relevant at higher photon transverse momentum (i.e. $p_\perp^{ll}$ in the dilepton case), where the coherent production mechanism is suppressed by the elastic ion form factor. At higher transverse momentum, the dominant emission process is due to incoherent inelastic emission from the nucleons in the ion. Therefore, to good approximation, one can follow a technique as in~\cite{ATLAS:2022srr} to model this process, namely by using the \texttt{SuperChic} distribution from dissociative lepton pair production in $pp$ collisions and setting the normalization in a a data--driven way; this is found to work rather well in the ATLAS analysis. A related excitation mechanism is for both ions to be excited by a single photon exchange in addition to the $\gamma\gamma$ initiated process, i.e. for the additional photon exchange not to be due to coherent emission from either ion. This is however suppressed for the same reason, being subleading in $Z$, and indeed as it is relevant at larger exchange virtualities, contributes in the region of smaller ion--ion impact parameters, which are strongly suppressed by the ion--ion survival factor. Other higher order QED processes that we will not consider in detail here are when additional photons are exchanged between the production process (i.e. the dilepton system in the case of lepton pair production) and the ions, as well as the production of additional low momenta electron pairs, i.e. so--called unitarising effects. These are discussed further in~\cite{Harland-Lang:2021ysd} and the references therein; while these will dominantly not lead to ion dissociation, they may effect the cross section predictions at the precision level.  A more detailed account of these is left to future work.
 
\begin{figure}
\begin{center}
\includegraphics[scale=0.62]{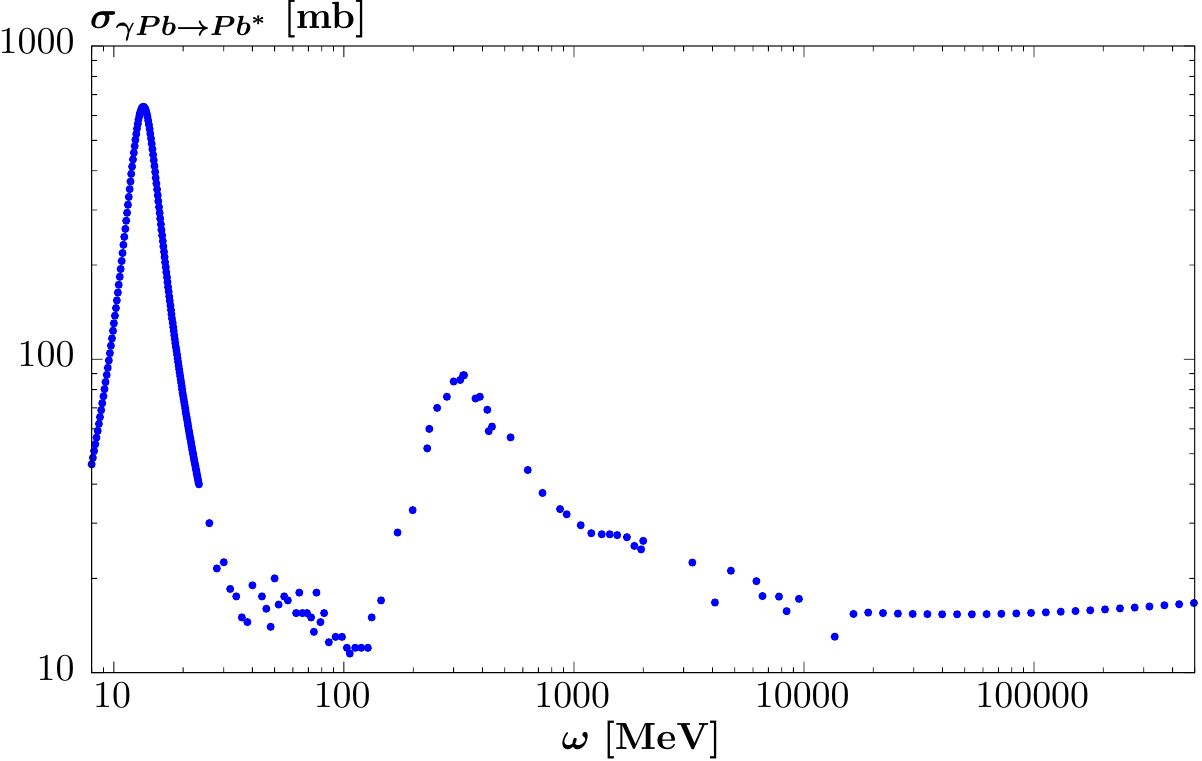}
\includegraphics[scale=0.62]{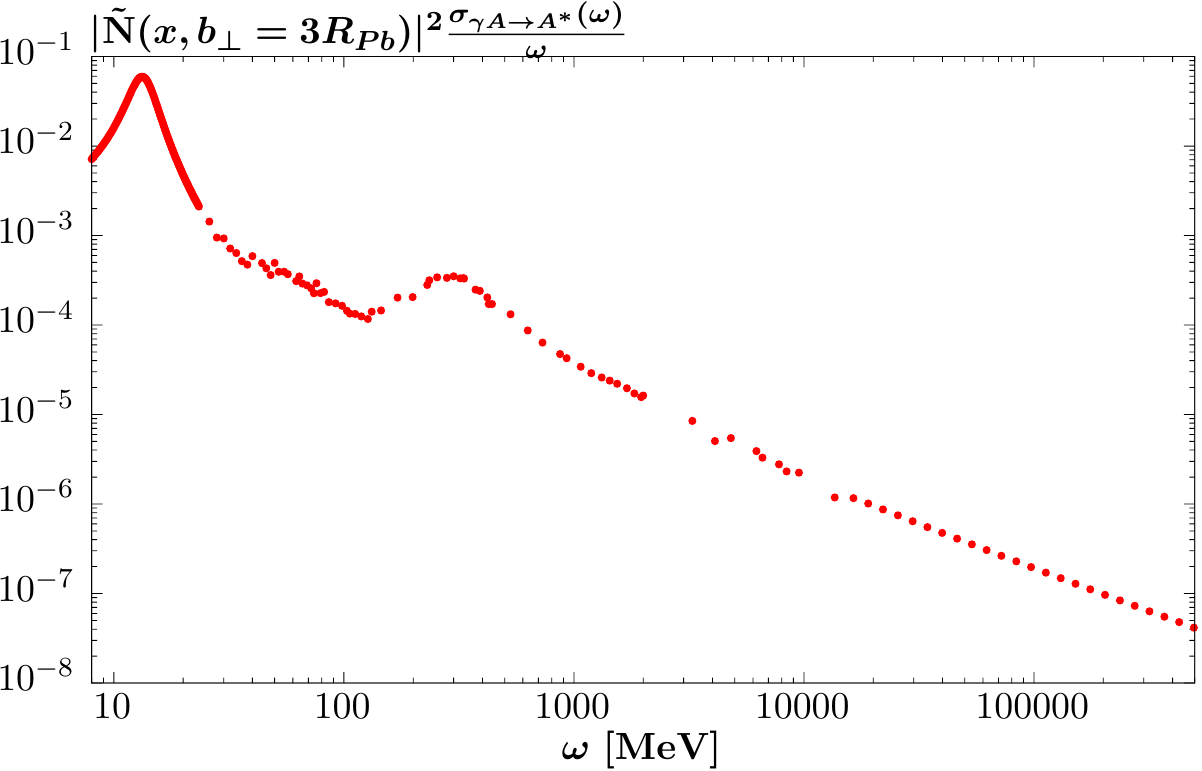}
\caption{(Left) Photon absorption cross section, $\sigma(\gamma A \to A^*)$, for lead ions, as a function of the photon energy $\omega$ in the dissociating ion rest frame. (Right) Photon absorption cross section weighted by corresponding flux factor as in \eqref{eq:p1} for $b_\perp = 3 R_A$, and $\sqrt{s}_{nn}=5.02$ TeV.}
\label{fig:siggam}
\end{center}
\end{figure}

 Returning to the case of mutual ion dissociation we will consider here, the nuclear excitation is assumed to occur independently of the $\gamma \gamma \to X$ production process, and so in impact parameter space we can write
 \be\label{eq:sigxx}
 {\rm d}\sigma_{X_1 X_2} = \int {\rm d}^2 b_{1\perp}\,{\rm d}^2  b_{2 \perp}\,  {\rm d}\sigma_{\rm S^2}\, P_{X_1 X_2}(s,b_\perp)\;,
 \ee
 where $P_{X_1 X_2}$ is the breakup probability, such that $X_{i} = 0,1,X$ corresponds to the the emission of 0 (i.e. no nuclear excitation) 1, or $X > 0$ neutrons emitted for each ion $i=1,2$. The probability $P_{X_1 X_2}(b_\perp)$ depends on the total ion--ion impact parameter $b_\perp = |  {\mathbf b}_{1\perp}+{\mathbf b}_{2\perp}|$ and factorizes into independent breakup probabilities for each ion, i.e.
 \be\label{eq:px1px2}
 P_{X_1 X_2}(b_\perp) = P_{X_1}(b_\perp)P_{X_2}(b_\perp)\;.
 \ee
 In the same way as  the $\gamma\gamma \to X$ cross section \eqref{eq:csepa} in the EPA,  formulated in impact parameter space, the lowest order breakup probability for each ion $i=1,2$ is then given by a convolution of the photon emission flux from the ion $j=2,1$ and the $\gamma A \to A^*$ cross section:
\be\label{eq:p1}
P^{1}_{Xn} (b_\perp) = \int\frac{ {\rm d} \omega}{\omega}\, |\tilde{\mathbf N}(x,b_{\perp})|^2  \sigma_{\gamma A \to A^*}(\omega)\;,
\ee
where $\omega$ is the photon energy in the $A$ rest frame, i.e. $\omega=xs/(2m_A)$ in the $s \gg m_A^2$ limit, which holds to very good approximation; the dissociation probability therefore depends on $s$, as well as the ion beam type, but we omit these arguments for brevity.
 The flux factor $|\tilde{N}|^2$ is built from exactly the same ingredients that enter the $\gamma\gamma \to X$ cross section \eqref{eq:csn}. To be precise it is given in terms of the Fourier transform
 \be
 \tilde{\mathbf N}(x,b_{\perp})\equiv\frac{1}{(2\pi)} \int {\rm d}^2 q_{\perp} {\mathbf N}(x,q_{\perp}) e^{-i {\mathbf q}_{\perp}\cdot  {\mathbf b}_{\perp}}\;,
 \ee
where ${\mathbf N}(x_i)$ is defined in \eqref{eq:nibf}. $\sigma_{\gamma A \to A^*}(\omega)$ is the photon--ion excitation cross section, which has been measured over a wide range of photon energies from fixed target ion scattering experiments. The corresponding data points, which we use to perform the integral \eqref{eq:p1} are show in Fig.~\ref{fig:siggam}; these are in many cases as implemented in the \texttt{Starlight} MC~\cite{Klein:2016yzr} (see also~\cite{Baltz:1996as,Broz:2019kpl}), with some exceptions that we will describe below. 

The cross section is shown in Fig.~\ref{fig:siggam} (left), while the result weighted by the flux factor as per \eqref{eq:p1} is shown in Fig.~\ref{fig:siggam} (right) for the, as we will see, representative value of $b_\perp = 3 R_A$. The largest peak in the $\omega \sim 10-20$ MeV region corresponds to the dominant GDR resonance, data for which is taken from~\cite{Veyssiere:1970ztg} for both the single and multiple neutron emission cases, fitted to a Lorentz shape. In the single neutron emission case the upper limit of the integral \eqref{eq:p1} is set to the upper limit of these data, at $\omega = 23.4$ MeV, where indeed the cross section is negligible. Above this energy, photon absorption will very dominantly lead to multiple neutron emission. In the $23.4 < \omega < 440$ MeV region we use the data from~\cite{Lepretre:1981tf,Carlos:1984lvc} on photonuclear scattering. In the $440 < \omega < 2000 $ MeV region we use the data from~\cite{Muccifora:1998ct} on photonuclear scattering; this is in contrast to~\cite{Klein:2016yzr} (see also~\cite{Baltz:1996as,Broz:2019kpl}) where older data~\cite{Armstrong:1971ns,Armstrong:1972sa} on $\gamma p$ and $\gamma n$ scattering were used, scaled to the nuclear case but with no shadowing applied. In the $2 < \omega < 16.4$ GeV region, we use data from~\cite{Caldwell:1973bu,Michalowski:1977eg} on $\gamma Pb$ and $\gamma Au$ scattering, suitably interleaved to cover the full energy region and with appropriate $A$ rescaling applied. In our MC implementation, for the GDR region data for both Au and Pb beams are implemented, with $A$ rescaling applied in other cases. For energies above this an effective per nucleon cross section is derived as described above (i.e. using a selection of Au and Pb data), which can be scaled to the appropriate ion $A$.

Finally, above 16.4 GeV there is some limited direct photon--ion data in the analysis of~\cite{Caldwell:1978ik}, in particular in the $\sim 40 - 80$ GeV region for Pb ions. To cover the entire energy region, including energies beyond this, we use the fact that the $\gamma$--nucleon cross section is observed to obey approximate Regge scaling as in~\cite{Donnachie:1992ny}. The high--energy data for this is limited to a handful of measurements at HERA, and for concreteness we take the ZEUS extraction~\cite{ZEUS:2001wan}. The $\gamma$--nucleon cross section is then suitably scaled by $A$, but with a nuclear shadowing factor of $0.65$ applied, in order to match the direct data from~\cite{Caldwell:1978ik} in the Pb case. 

The contribution from this region, while naively suppressed by $\sim 1/\omega$, as evidenced in Fig.~\ref{fig:siggam} (right), is in fact in principle rather significant, due to the large photon energies available. In particular, from the considerations in Section~\ref{sec:key} the photon $x$ is cutoff at roughly $x_{\rm max} \sim  1/(R_A m_A)$, and hence the photon $\omega_{\rm max} \sim s/(m_A^2 R_A) $. At the LHC this is of order 850 TeV, which corresponds to a $\gamma n$ c.m.s. energy that is roughly an order of magnitude larger than that  probed at HERA, and many orders of magnitude higher than the highest energy direct data on photon--ion absorption. To give an estimate, we can assume for simplicity that the flux term in \eqref{eq:p1} is constant up to the cutoff $x_{\rm max}$, and that the $\gamma A$ cross section is constant with energy, which 
is roughly consistent with the Regge parameterisation. We then have
\be\label{eq:omsig}
 \int_{\omega > \omega_0}\frac{ {\rm d} \omega}{\omega} \sigma_{\gamma A \to A^*}(\omega) \sim \sigma_{\gamma A \to A^*}^{\omega_0} \ln\left(\frac{\omega_{\rm max}}{\omega_0}\right)\sim 140 \,{\rm mb}\;,
\ee
if we take the measured value from~\cite{Caldwell:1978ik} at $\omega_0=80$ GeV. The (in theory dominant) contribution from the GDR region can be estimated using the TKR sum rule (see e.g.~\cite{Baur:1986uso})
\be
 \int_{\rm GDR}\frac{ {\rm d} \omega}{\omega} \sigma_{\gamma A \to A^*}(\omega) \sim \frac{1}{E_{\rm GDR}}\frac{60 NZ}{A} \,{\rm MeV}\,{\rm mb}\sim 220 \,{\rm mb}\;,
 \ee
for the Pb case. At the LHC, a more precise numerical evaluation gives 260 mb for the integrated single neutron emission cross section (equal to the total GDR contribution to reasonable approximation) and 177 mb for \eqref{eq:omsig} at a representative value of  $b_\perp = 3 R_A$,  consistent with these rough expectations. Including the contribution below 80 GeV and above the GDR region, we find that \eqref{eq:omsig} accounts for roughly 25\% of the total contribution to \eqref{eq:p1} at the LHC, again for $b_\perp = 3 R_A$. Given the lack of direct data in this region, we can conservatively assign an uncertainty of this order in the corresponding neutron tagged cross section. We note that a useful way to  bypass this uncertainty source is to consider single neutron tag data (i.e. $0n1n$ and $1n1n$), where the contribution from this region is negligible.

Indeed, at these high energies, the photon--ion interaction cannot be viewed as a purely electromagnetic one. In particular, according to the rather well established vector meson dominance model~\cite{Sakurai:1969ss,Sakurai:1960ju,Bauer:1977iq} we can view the photon as a superposition of light vector mesons, which then undergoes a hadronic (dominantly inelastic) interaction with a nucleon within the ion. 
Indeed, for sufficiently high photon energies the $\gamma n$ system will become relatively more central in rapidity, while the underlying $\gamma n$ interaction will be a relatively high multiplicity inelastic hadronic event. Hence it is arguably possible that the produced neutrons may not be detected in the ZDCs or, more significantly, whether some of the  products of the inelastic $\gamma n$ interaction will  be seen centrally and hence fail the experimental veto on additional particle production in the central detector. Given it has been observed in e.g. the ATLAS data on muon pair production~\cite{ATLAS:2020epq} in PbPb UPCs that the \texttt{Starlight}~\cite{Klein:2016yzr} predictions (which include this high energy contribution up to the kinematic limit) tend to overshoot the observed $0nXn$ and $XnXn$ fraction, we will to be precise cut the cross section off at $\omega = 500$ GeV, which corresponds to $|y_{\gamma n}| \sim 5$ for PbPb collisions at $\sqrt{s}_{nn} = 5.02$ GeV. As particle production will in general occur at rapidities lower than this, we may expect this to fail the experimental veto, although a precise evaluation would require that we account for the particle multiplicity distribution and the particular experimental cuts; for e.g. the ATLAS analyses~\cite{ATLAS:2022srr,ATLAS:2020epq} we will consider in the following section, the relevant requirements are for no additional tracks with $p_\perp > 100$ -- 200 MeV out to $|\eta| = 2.5-3.75$. Even absent this we can view this as an effective cutoff, driven by the comparison to data, and given the uncertainties in the calculation discussed above. For $b_\perp = 3 R_A$ this removes roughly 15\% of the contribution. As we will see when comparing to the ATLAS measurement of dilepton production in UPCs, it may be that a more stringent cut is required to match the data.

\begin{figure}
\begin{center}
\includegraphics[scale=0.62]{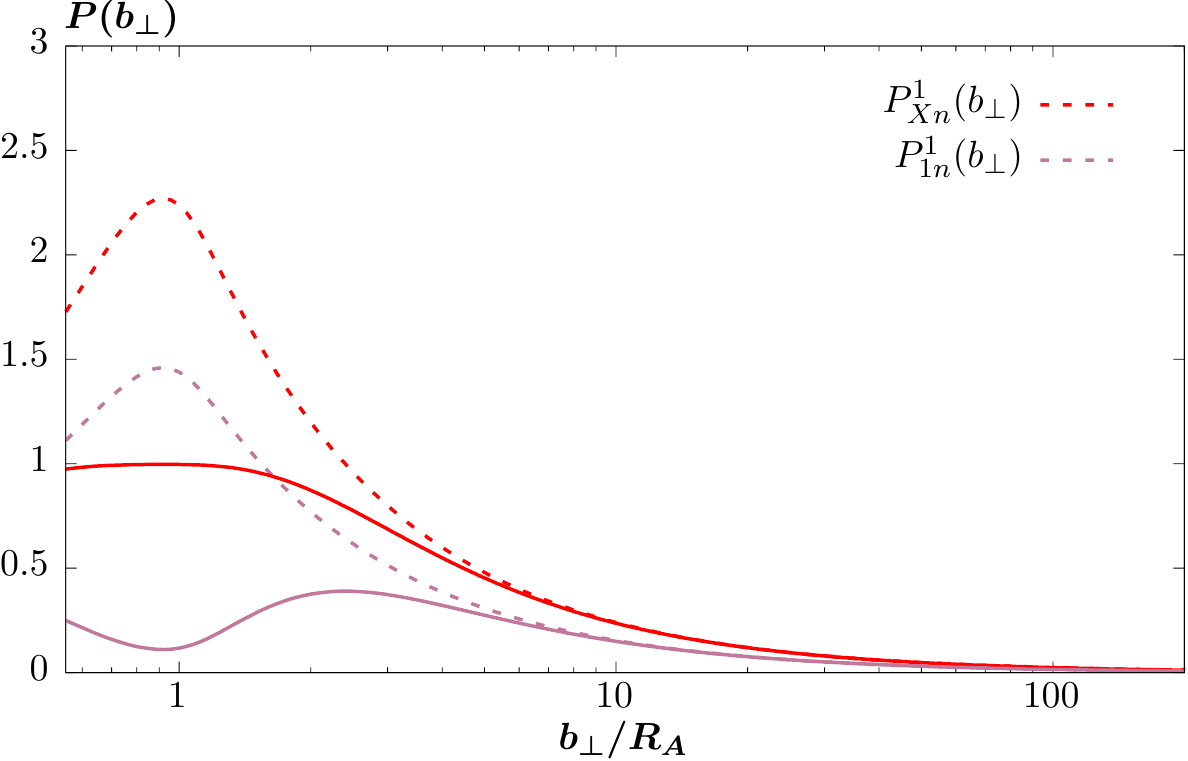}
\caption{Lowest order breakup probabilities for PbPb collisions at $\sqrt{s_{nn}}=5.02$ TeV, for single and multiple neutron emission, as defined in \eqref{eq:p1}, given by the dashed curves. Also shown in solid is the result applying the unitarising corrections discussed in the text.}
\label{fig:PXnounit}
\end{center}
\end{figure}

The resulting lowest order breakup probabilities, for single and multiple neutron emission, are shown in Fig.~\ref{fig:PXnounit} by the dashed lines. The multiple neutron emission probability is  moderately larger than the single emission probability, as expected from the discussion above. However, we can also see that at smaller impact parameters, the probability increases, and in both cases rises above unity. This is driven by the flux in \eqref{eq:p1} which one can readily show scales as
\be\label{eq:napprox}
|\tilde{\mathbf N}(x,b_{\perp})|^2 \approx \frac{Z^2\alpha}{\pi^2} \frac{1}{b_\perp^2}\;,
\ee
for the dominant region of $x$ (i.e. $x \ll 1$) that contributes to \eqref{eq:p1} and for $b_\perp > R_A$ such that the ion can be treated as a point--like charge. That is the flux is peaked towards low $b_\perp$, until we reach $b_\perp < R_A$ where the extended nature of the ion comes into play and the flux becomes suppressed by the ion form factor (although the contribution from this region is in any case negligible once one accounts for the ion--ion survival factor).

This indicates an inadequacy in the lowest order (in $Z^2\alpha$) perturbative approximation for the ion excitation process, in particular given the $Z^2$ enhanced flux for photon emission from the spectator ion. To account for this, we follow the approach described in~\cite{Baltz:2009jk}, namely assuming each ion excitation process happens independently we have that the number of excitations follows a Poissonian probability, such that
\begin{align}\nonumber
P_{0n} (b_\perp)  &= \exp(-P^1_{Xn} (b_\perp))\;,\\ \nonumber
P_{1n} (b_\perp)  &= P^1_{1n}\exp(-P^1_{Xn} (b_\perp))\;,\\ \label{eq:pXs}
P_{Xn} (b_\perp)  &= 1- \exp(-P^1_{Xn} (b_\perp))\;.
\end{align}
The impact of this unitarising is shown in Fig.~\ref{fig:PXnounit}, where we can see this by construction gives dissociation probabilities that never exceed unity. While the dominant effect of this is in fact below the $b_\perp \sim 2 R_A$ region, which is in any case removed once the ion--ion survival factor is included, it is not negligible; it reduces the $XnXn$ cross section by a factor of $\sim 1.5-2$, depending on the precise kinematics.

\begin{figure}
\begin{center}
\includegraphics[scale=0.62]{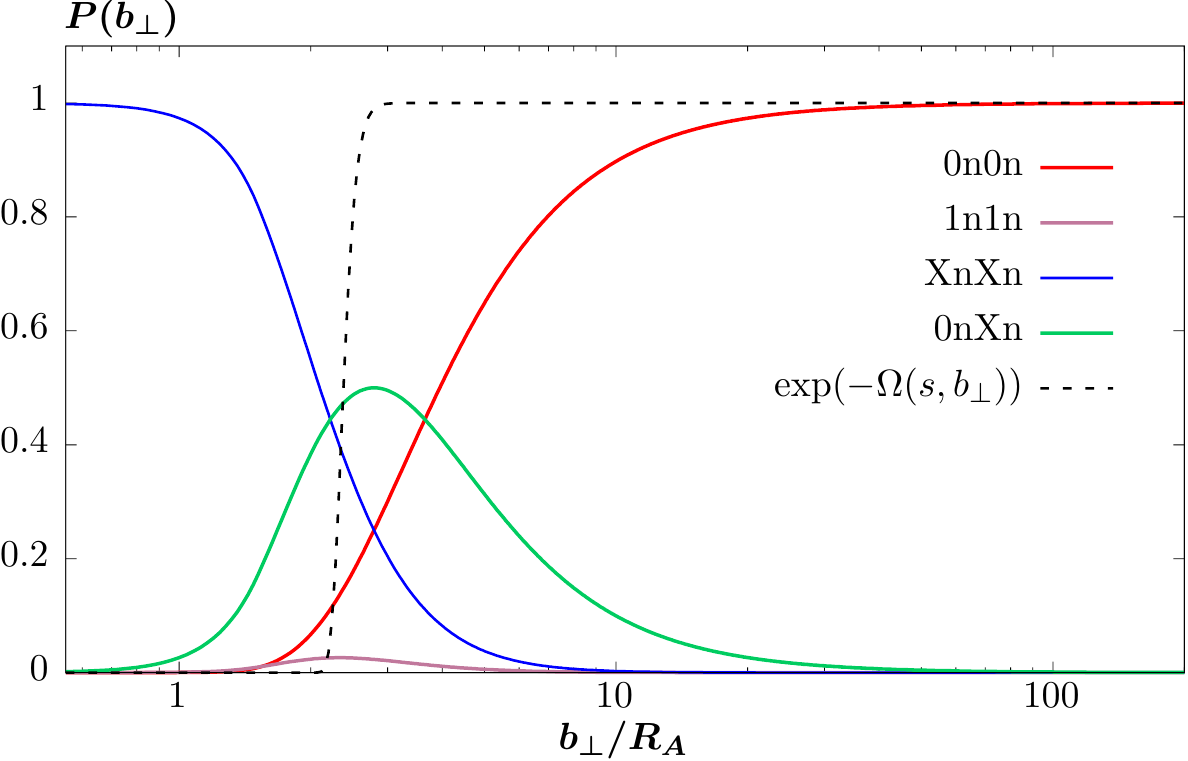}
\includegraphics[scale=0.62]{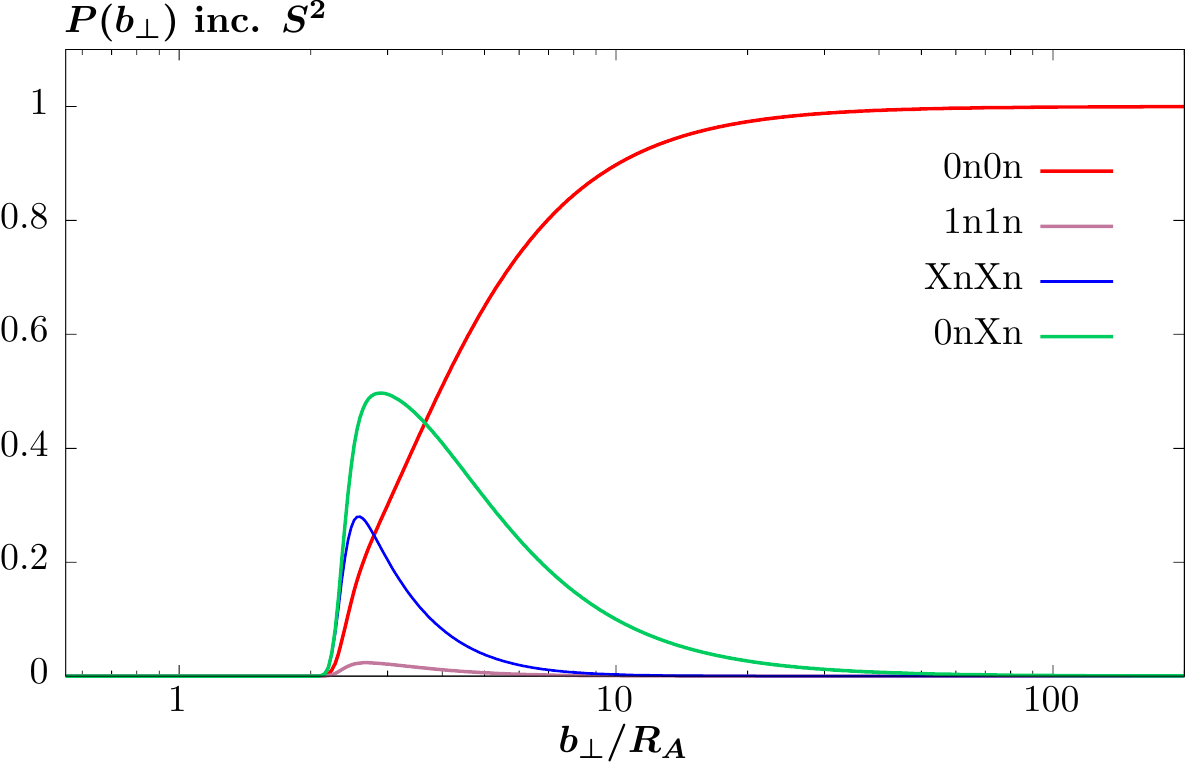}
\caption{(Left) Breakup probabilities for single and multiple neutron emission, as defined in \eqref{eq:pXs} and \eqref{eq:px1px2}, for PbPb collisions at $\sqrt{s_{nn}}=5.02$ TeV. The probability that no inelastic ion--ion scattering occurs, as introduced in \eqref{eq:opac}, is indicated by the dashed black curve. (Right) As in the left figure, but now multiplied by the no inelastic ion--ion scattering probability, i.e. including the survival factor.}
\label{fig:PX}
\end{center}
\end{figure}

The corresponding dissociation probabilities \eqref{eq:px1px2} are shown in Fig.~\ref{fig:PX}. In the left plot the results prior to multiplying by the probability $\exp(-\Omega_{A_1 A_2}(s,b_\perp))$ of no ion--ion inelastic scattering are shown for reference, and in the right plot this is accounted for. As expected from the discussion above, the probability for ion dissociation is peaked towards $b_\perp \sim 2 R_A$ by the scaling of the flux \eqref{eq:napprox}, before being cut off by the no ion--ion inelastic scattering probability, which rapidly approaches zero below $b_\perp \lesssim 2 R_A$. Conversely, if no ion dissociation is required, as in the $0n0n$ case, than the probability become increasingly close to unity as $b_\perp$ increases. For the mixed $0nXn$ case the peaking towards $b_\perp \sim 2 R_A$ is again present, although this is less strong than for $XnXn$; the $0n1n$ result is not shown for clarity, though it follows a similar trend.

So far, we have worked purely in impact parameter space, however for a full treatment, and in particular to account for the process and kinematic dependence of the ion dissociation probability, we must translate these to transverse momentum space. To do this, we simply replace 
\be\label{eq:gamrep}
\Gamma_{A_1A_2}(s,b_\perp)^{1/2} \to \left[ \Gamma_{A_1A_2}(s,b_\perp) P_{X_1 X_2}(b_\perp)\right]^{1/2}\;,
\ee
and then use this as in \eqref{eq:tqts2} to get the corresponding amplitude in transverse momentum space and hence cross section. More precisely, from \eqref{eq:pmc1} we are interested in the integral
\begin{align}\nonumber
&\int {\rm d}^2 b_{\perp}\,e^{i {\mathbf k}_{\perp}\cdot {\mathbf b}_{\perp}} \left[ \Gamma_{A_1A_2}(s,b_\perp) P_{X_1 X_2}(b_\perp)\right]^{1/2}\;,\\ \label{eq:pXbess}
& = 2\pi \int {\rm d}b_\perp \, b_\perp J_0(b_\perp k_\perp) \left[ \Gamma_{A_1A_2}(s,b_\perp) P_{X_1 X_2}(b_\perp)\right]^{1/2}\;,
\end{align}
where $J_0$ is a Bessel function of the first kind. Making use of \eqref{eq:napprox} we have
\be
P^{1}_{Xn} (b_\perp) \approx \frac{A_{Xn}}{b_\perp^2}\;,
\ee
where
\be
A_{Xn} =   \frac{Z^2\alpha}{\pi^2} \int\frac{ {\rm d} \omega}{\omega}\,  \sigma_{\gamma A \to A^*}(\omega)\;.
\ee
The large $b_\perp$ limit of \eqref{eq:pXbess}, for which we have $\Gamma_{A_1A_2}(b_\perp) \sim 1$, is therefore driven by the integrand
\be
I_{X_1 X_2}(b_\perp) \equiv b_\perp J_0(b_\perp k_\perp) P_{X_1 X_2}(b_\perp)^{1/2}\;,
\ee
where in this limit we have
\begin{align}\nonumber
P_{0n} (b_\perp)  &\approx 1-\frac{A_{Xn}}{b_\perp^2}\;,\\ \nonumber
P_{1n} (b_\perp)  &\approx \frac{A_{1n}}{b_\perp^2}\left(1-\frac{A_{Xn}}{b_\perp^2}\right)\;,\\ \label{eq:pXslbt}
P_{Xn} (b_\perp)  &\approx \frac{A_{Xn}}{b_\perp^2}\;,
\end{align}
such that 
\be\label{eq:11scal}
I_{1n 1n},\,I_{1n Xn},\,I_{Xn Xn} \sim \frac{J_0(b_\perp k_\perp)}{b_\perp}\;,
\ee
which is numerically rapidly converging\footnote{Strictly speaking this result relies on  the $x b_\perp m_A \ll 1$ limit being true, and hence is not valid at very large values of $b_\perp$. However this occurs in a region where the integrand is already numerically negligible and moreover for large $x b_\perp m_A \gg 1$ scaling is in fact more strongly (exponentially) suppressed by $b_\perp$, such that the corresponding integral \eqref{eq:pXbess} is certainly convergent.}. For the same reason, once we appropriately use \eqref{eq:pmc2} rather than \eqref{eq:pmc1} then the $0n0n$ results in an integral that has the same convergence. For the remaining cases however we have
\be
I_{0n 1n},\,I_{0n Xn} \sim J_0(b_\perp k_\perp)\;,
\ee
and the numerical convergence of the integral (though it is certainly finite) is more problematic. To resolve this, we can consider the integrands
\be
\left[I_{0n 1n}- \frac{A_{1n}^{1/2}}{b_\perp}\right] + \frac{A_{1n}^{1/2}}{b_\perp}\;.
\ee
The first (square bracketed) term now to good approximation scales as $\sim J_0(b_\perp k_\perp)/b_\perp^2$ and is hence numerically convergent, while the latter can be evaluated as
\begin{align}\nonumber
&2 \pi A_{1n}^{1/2}\int {\rm d} b_\perp\,J_0(b_\perp k_\perp)   \Gamma_{A_1A_2}(s,b_\perp)^{1/2}\;,\\ \label{eq:1nap}
&= 2 \pi A_{1n}^{1/2}\int {\rm d} b_\perp\,J_0(b_\perp k_\perp) \left[ \Gamma_{A_1A_2}(s,b_\perp)^{1/2}-1\right]+\frac{2 \pi A_{1n}^{1/2}}{k_\perp}\;,
\end{align}
where we have used 
\be
\int_0^\infty {\rm d} t \, J_0(t) = 1\;,
\ee
while we can see that the integrand of the first term in \eqref{eq:1nap} only has support for $b_\perp \sim 2 R_A$ and hence can readily be evaluated. The $0nXn$ case can be evaluated in a similar way, or alternatively by rearranging
\be
{\rm d} \sigma_{\rm inc.} = {\rm d} \sigma_{0n0n} +  {\rm d} \sigma_{0nXn} +{\rm d} \sigma_{XnXn}\;,
\ee
where `inc.' denotes the inclusive case, with no ion dissociation requirement applied, and for the $0nXn$ it is always implied that either ion can dissociate (but not both).

Having established the framework to account for ion dissociation in  transverse momentum space, in the following sections we will now consider results of this and compare them to data. We note that, as discussed at the end of Section~\ref{sec:surv} for the case of the ion--ion survival factor, the impact of the ion dissociation requirement (i.e. the predicted cross section for ion dissociation) will depend on the  kinematics of the central system as well as the corresponding process under consideration. This is clear from Fig.~\ref{fig:PX}, where we can see that the impact parameter dependence of the dissociation probability is different depending on the particular neutron tag. This will therefore lead to a non--trivial dependence in the Fourier conjugate transverse momentum space, as we will see.

\section{Results}\label{sec:results}

\subsection{Comparison to ATLAS data}

\begin{figure}[t]
\begin{center}
\includegraphics[scale=0.62]{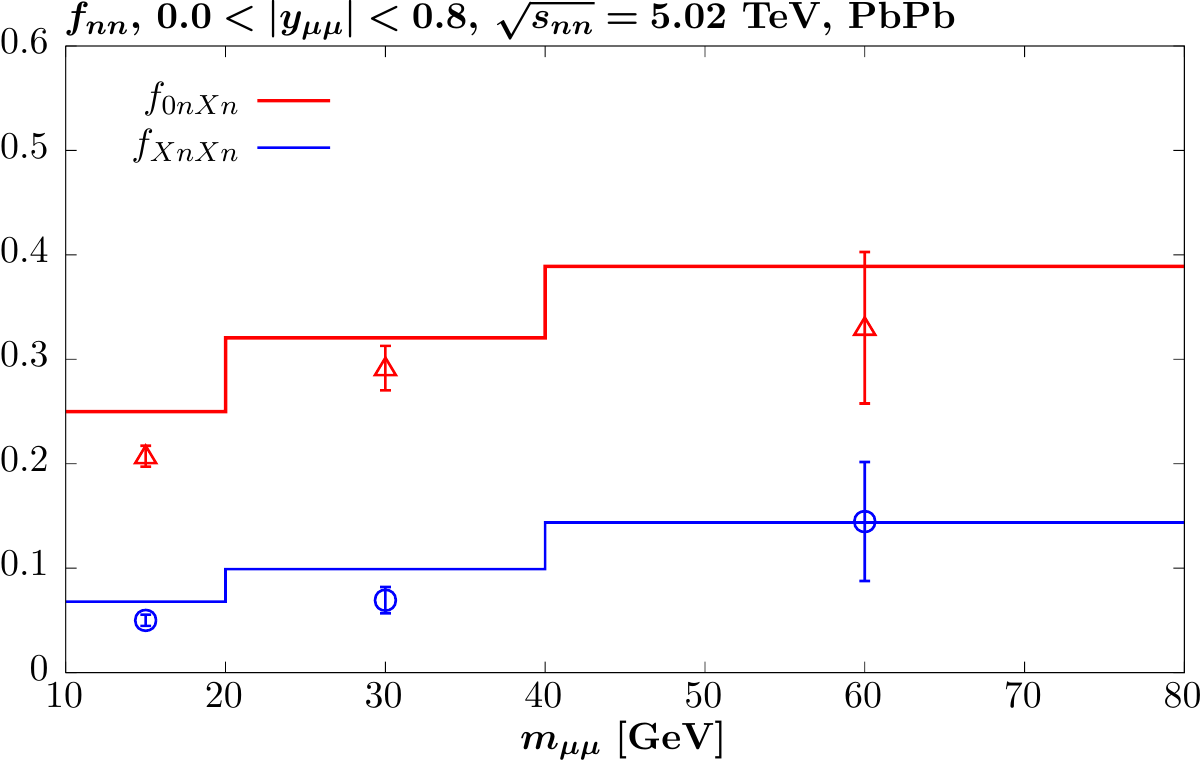}
\includegraphics[scale=0.62]{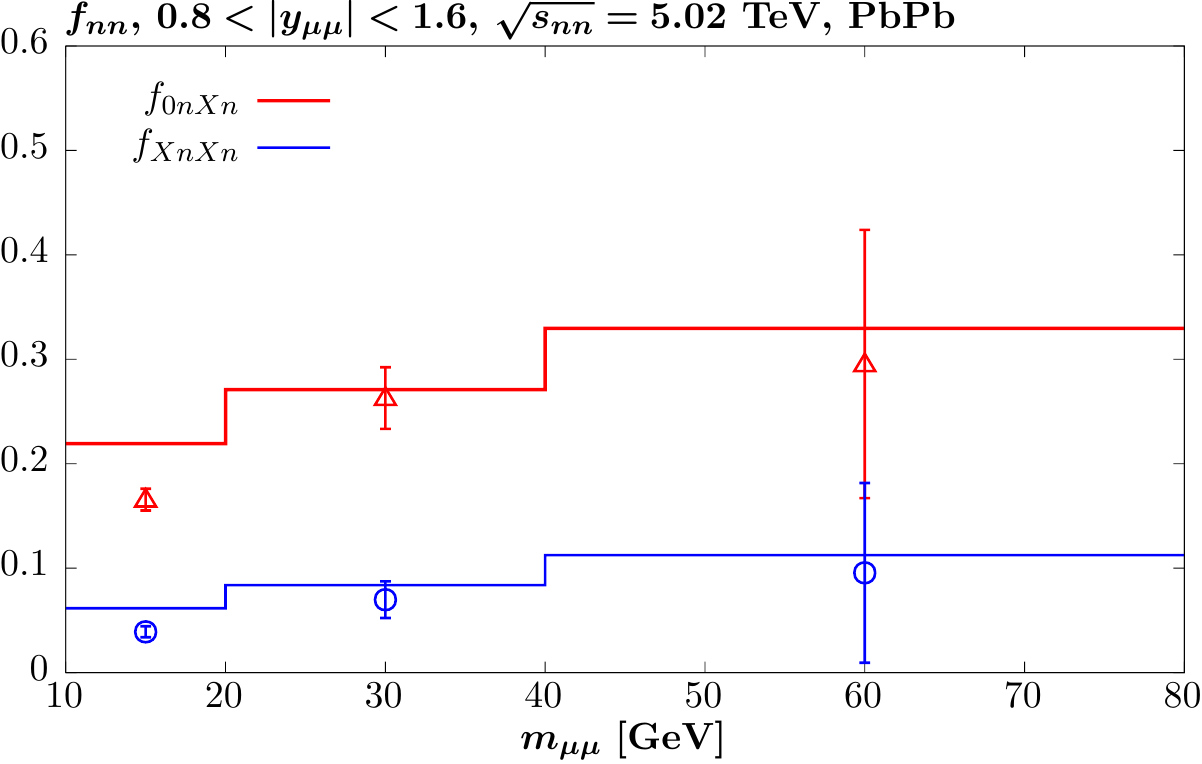}
\includegraphics[scale=0.62]{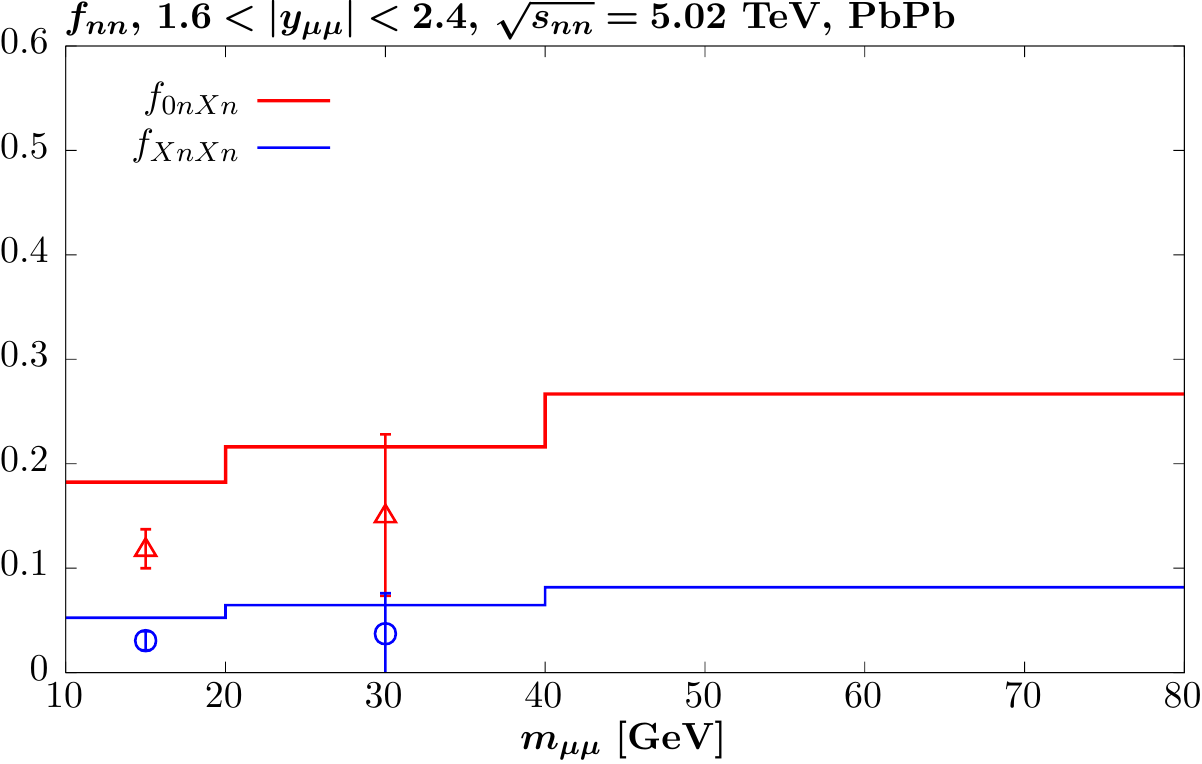}
\caption{Comparison of \texttt{SuperChic 4.2} predictions  to ATLAS data~\cite{ATLAS:2020epq} on ultraperipheral muon pair production in PbPb collisions at $\sqrt{s_{nn}}=5.02$ TeV as a function of the dimuon invariant mass and for different dimuon rapidity regions. Results for the ratio of the $Xn0n$ and $XnXn$ cross sections to the inclusive UPC case (with respect ion dissociation) are shown. The muons are required to have $p_{\perp,\mu}>4$ GeV, $|\eta_\mu|<2.4$, $m_{\mu\mu} > 10$ GeV and $p_{\perp,\mu\mu}< 2$ GeV. Data errors correspond to systematic and statistical added in quadrature.}
\label{fig:ATLASrap}
\end{center}
\end{figure}

\begin{figure}[t]
\begin{center}
\includegraphics[scale=0.62]{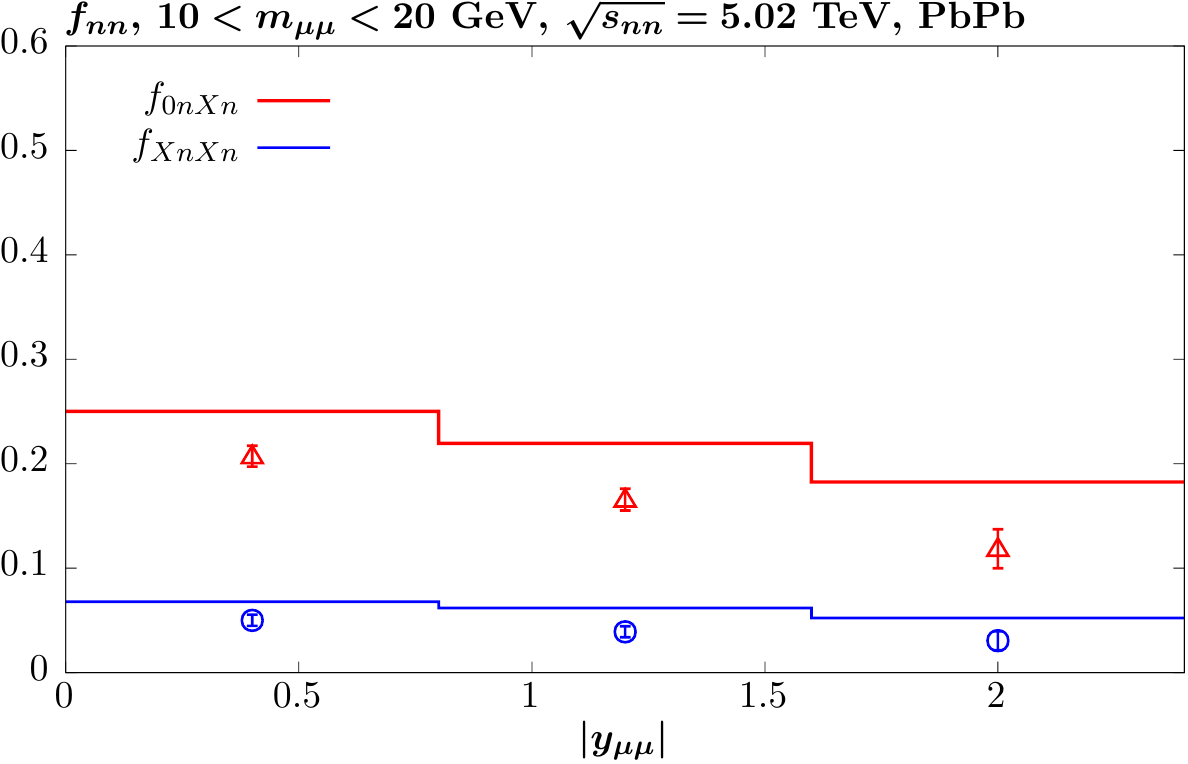}
\includegraphics[scale=0.62]{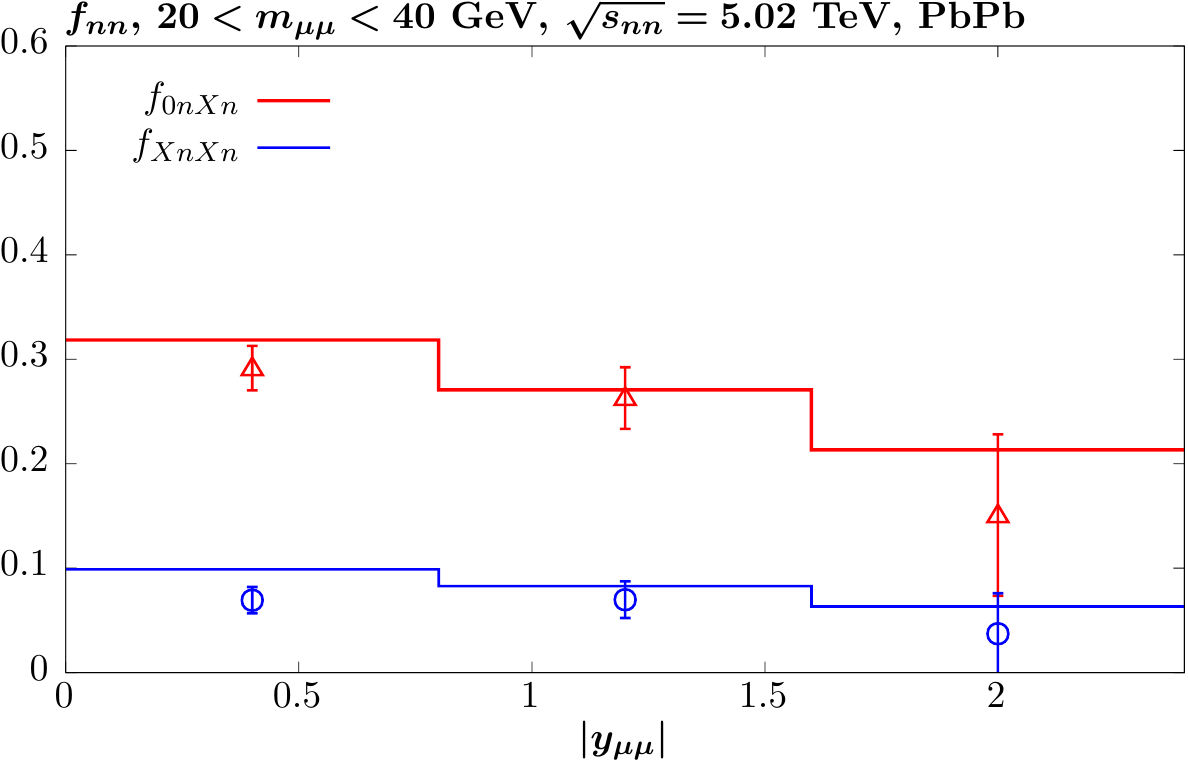}
\includegraphics[scale=0.62]{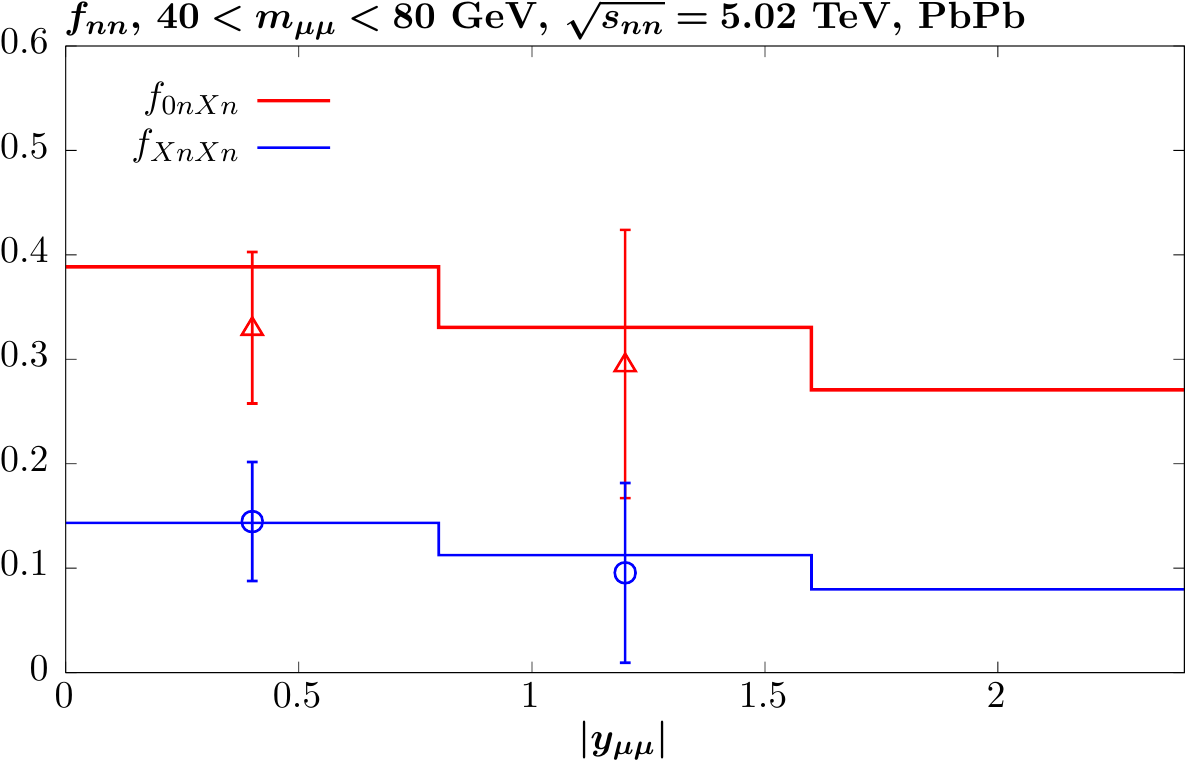}
\caption{As in Fig.~\ref{fig:ATLASrap} but now shown as a function of the dimuon rapidity, and for different dimuon invariant mass regions.}
\label{fig:ATLASm}
\end{center}
\end{figure}

In this section we will compare our predictions to ATLAS measurements of muon~\cite{ATLAS:2020epq} and electron~\cite{ATLAS:2022srr} pair production in ultraperipheral PbPb collisions at $\sqrt{s_{nn}}=5.02$ TeV. We will focus throughout on those observables that are presented explicitly with a neutron tag imposed in the ZDCs, although in~\cite{ATLAS:2020epq,ATLAS:2022srr} the inclusive data are compared to \texttt{SuperChic} predictions, with the agreement in general found to be good. We note that by `tag' it is always implied that neutrons are either required or vetoed on, while by inclusive we mean that both cases are included, i.e. no ZDC requirement is made, although we still require the collisions to be ultraperipheral, with no colour flow between the ions. 

In the ATLAS case, three event categories are considered, namely the case where no neutrons are registered in the ZDCs, where $X>0$ neutrons are registered on one side but none on the other, and where $X>0$ neutrons are registered on both sides, denoted by $0n0n$, $0nXn$ and $XnXn$, respectively. The cross section fractions $f_{in jn}$ are then defined relative to the inclusive case, such that
\be\label{eq:fsum}
f_{0n0n} + f_{0nXn} + f_{XnXn}=1\;.
\ee
We begin by comparing to the data on muon pair production~\cite{ATLAS:2020epq}. In Fig.~\ref{fig:ATLASrap} we show the $0nXn$ and $XnXn$ fractions as a function of the dimuon invariant mass and for different dimuon rapidity regions; the remaining $0n0n$ fraction is then found using \eqref{eq:fsum} and so is not independent of these. In terms of the broad trend, we can see that both of these fractions are predicted to increase with $m_{\mu\mu}$, which is as we would expect. In particular, from \eqref{eq:qi} we can see that the average initial--state photon transverse momentum will increase with the photon momentum fraction $x_i \propto m_{\mu\mu}$. In impact parameter space this corresponds to smaller $b_\perp$ values, and we can see from Fig.~\ref{fig:PX} that it is precisely the $0nXn$ and $XnXn$ cases that are enhanced in this region, with the $0n0n$, conversely, suppressed. This trend is clearly observed in the data, even within the relatively large uncertainties at larger $m_{\mu\mu}$. In more detail, however, we can see that there is a distinct tendency to overshoot the data in the lowest mass bin (and at central rapidities, in the second mass bin). That is, too much ion dissociation is predicted, for all dimuon rapidities. 

In Fig.~\ref{fig:ATLASm} we show a comparison to the same dataset, but now presented differentially in the dimuon rapidity, for different dimuon invariant mass regions. We can see that the  $0nXn$ and $XnXn$ fractions are predicted to increase with rapidity, which is again driven by the changing photon $q_\perp$ dependence in the production amplitude due to \eqref{eq:qi}, and the dependence of the photon momentum fractions on the dimuon rapidity. In this case, forward rapidity corresponds to an increased photon $x_i$ on one side, but a decrease on the other, and so it is not immediately obvious that the overall trend should be for larger average photon transverse momenta, and hence smaller impact parameters $b_\perp$. This is however the basic trend predicted by the full numerical treatment, and indeed similar behaviour is predicted in earlier studies of photon--initiated production in heavy ion and $pp$ collisions, see e.g.~\cite{Harland-Lang:2018iur,Harland-Lang:2020veo}. We can see from the plots that this trend is indeed observed in the data, again supporting the theoretical framework presented here. On the other hand, the same overshoot in the lowest $m_{\mu\mu}$ bin is clear. 

\begin{figure}[t]
\begin{center}
\includegraphics[scale=0.62]{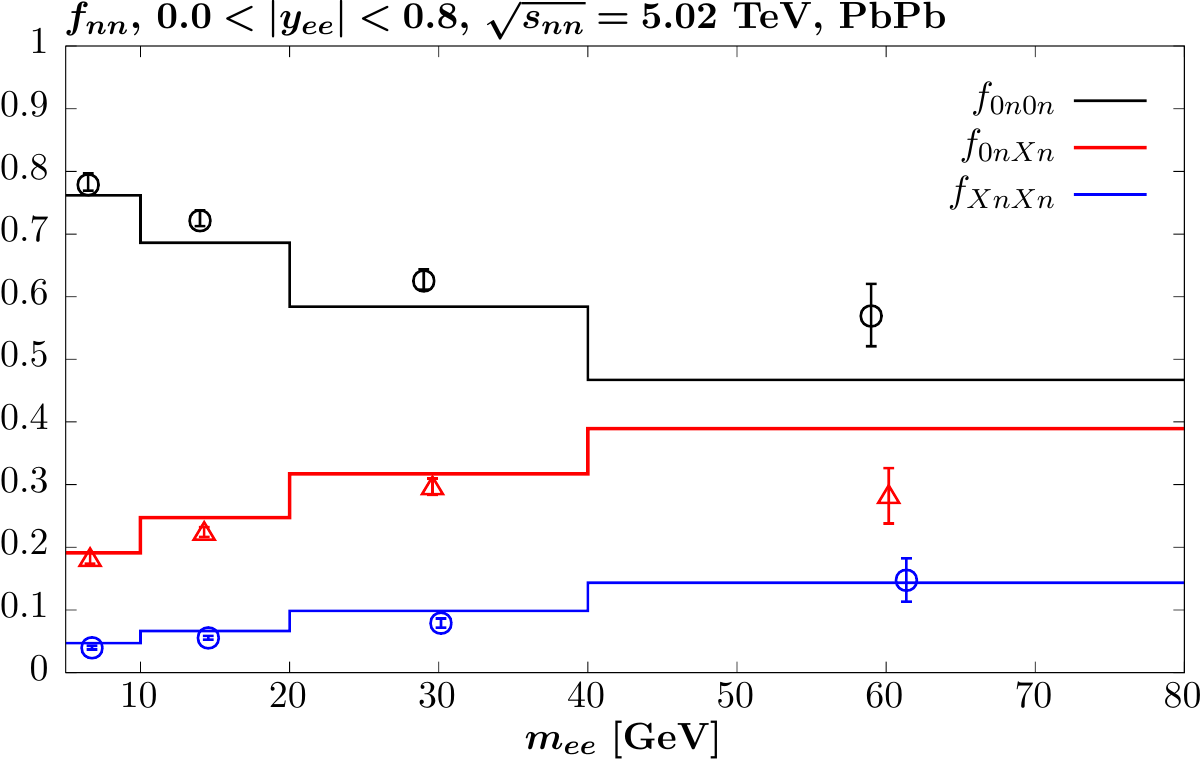}
\includegraphics[scale=0.62]{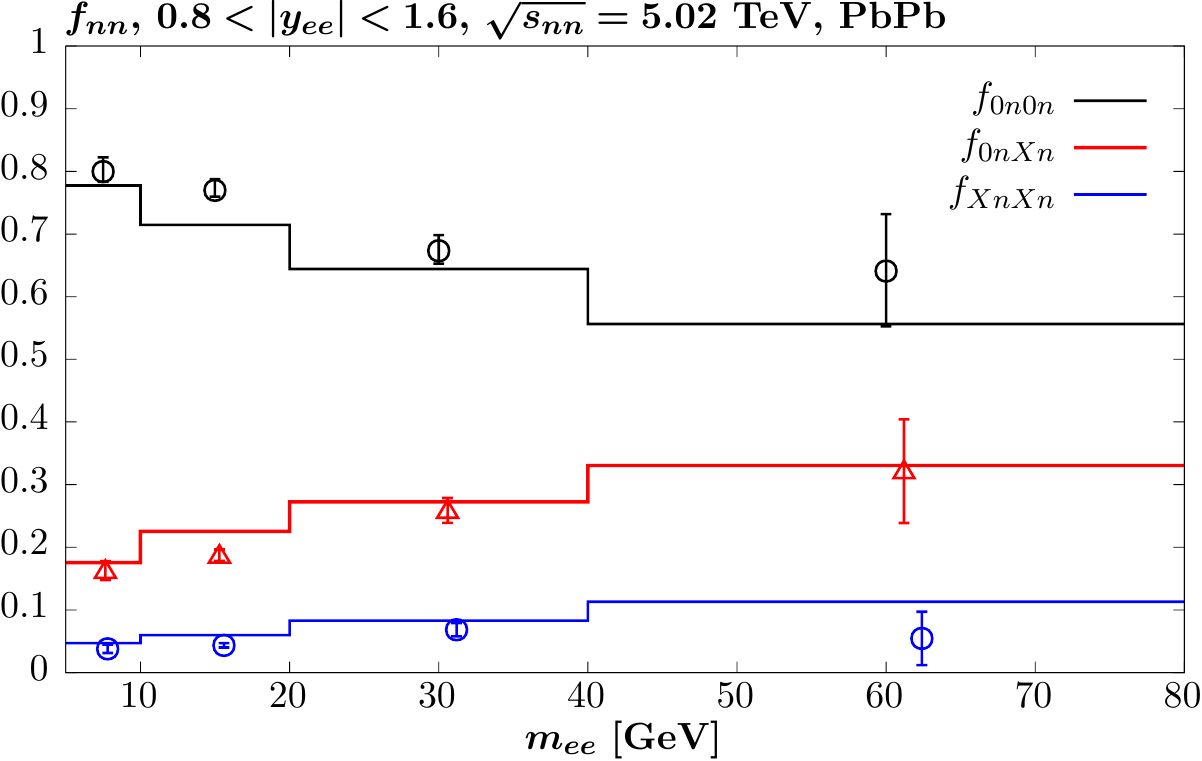}
\includegraphics[scale=0.62]{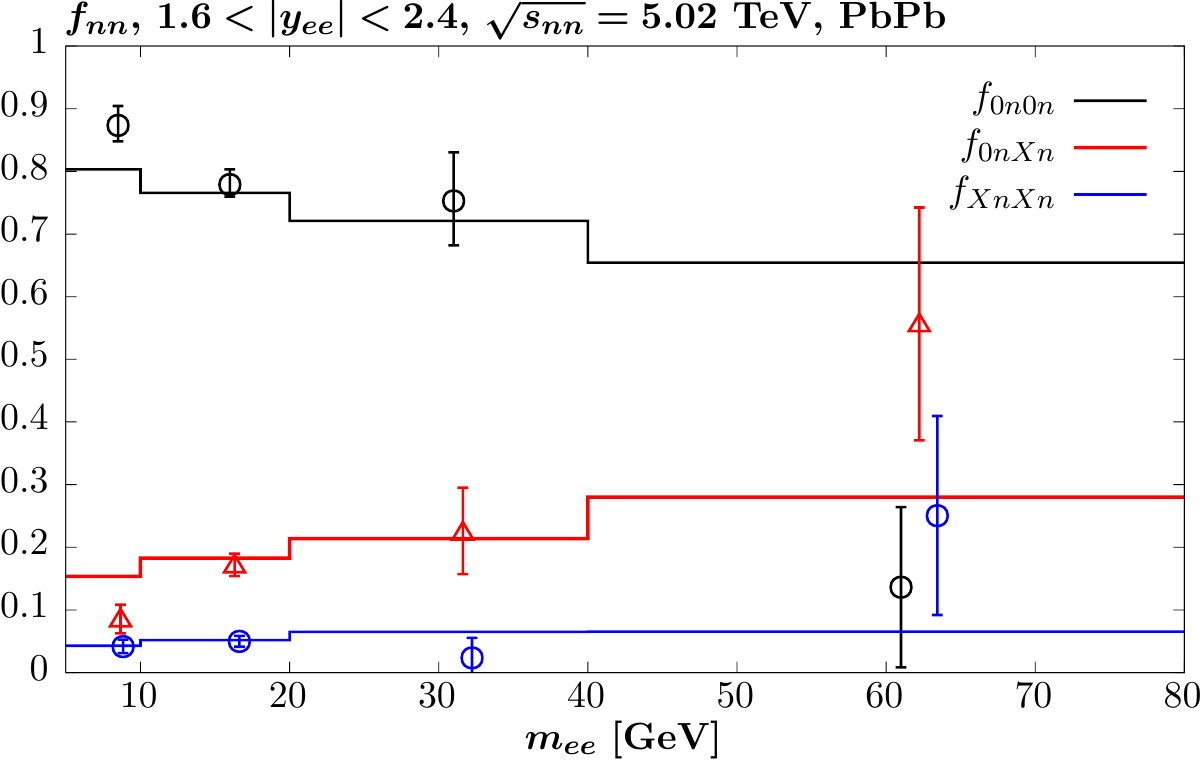}
\caption{Comparison of \texttt{SuperChic 4.2} predictions  to ATLAS data~\cite{ATLAS:2022srr} on ultraperipheral electron pair production in PbPb collisions at $\sqrt{s_{nn}}=5.02$ TeV as a function of the dimuon invariant mass and for different dimuon rapidity regions. Results for the ratio of the $0n0n$, $Xn0n$ and $XnXn$ cross sections to the inclusive UPC case (with respect ion dissociation) are shown. The electrons are required to have $p_{\perp,e}>2.5$ GeV, $|\eta_e|<2.4$, $m_{ee} > 5$ GeV and $p_{\perp,\mu\mu}< 2$ GeV. Data errors correspond to systematic and statistical added in quadrature.}
\label{fig:ATLASdiel}
\end{center}
\end{figure}

We next consider a comparison to the ATLAS data on electron pair production~\cite{ATLAS:2022srr}. In Fig.~\ref{fig:ATLASdiel} we show the $0n0n$, $0nXn$ and $XnXn$ fractions as a function of the dielectron invariant mass, for different dielectron rapidity regions. As data for the three cases are provided by ATLAS, we present comparisons for all of these, however as noted above only two of the three are independent, i.e. the sum in any given bin is by construction unity. We note that the event selection, given in the figure captions, is rather similar between this and the muon measurement, with in this case somewhat lower invariant masses being probed. The same overall trend as predicted above is again seen with respect to the pair invariant mass, $m_{ee}$, i.e. for the $0nXn$ and $XnXn$ fractions to increase, and the $0n0n$ to decrease, as $m_{ee}$ increases. This is again clearly observed in the data, and in general the level of agreement between data and theory is good. The most visible differences are in the forward rapidity, $1.8 < |y_{ee}|<2.4$, bin, both at low and high masses. In the high mass bin, however, the data errors are rather large and certainly the rather extreme suppression in the $0n0n$ case is not seen in the other rapidity bins.

\begin{figure}
\begin{center}
\includegraphics[scale=0.6]{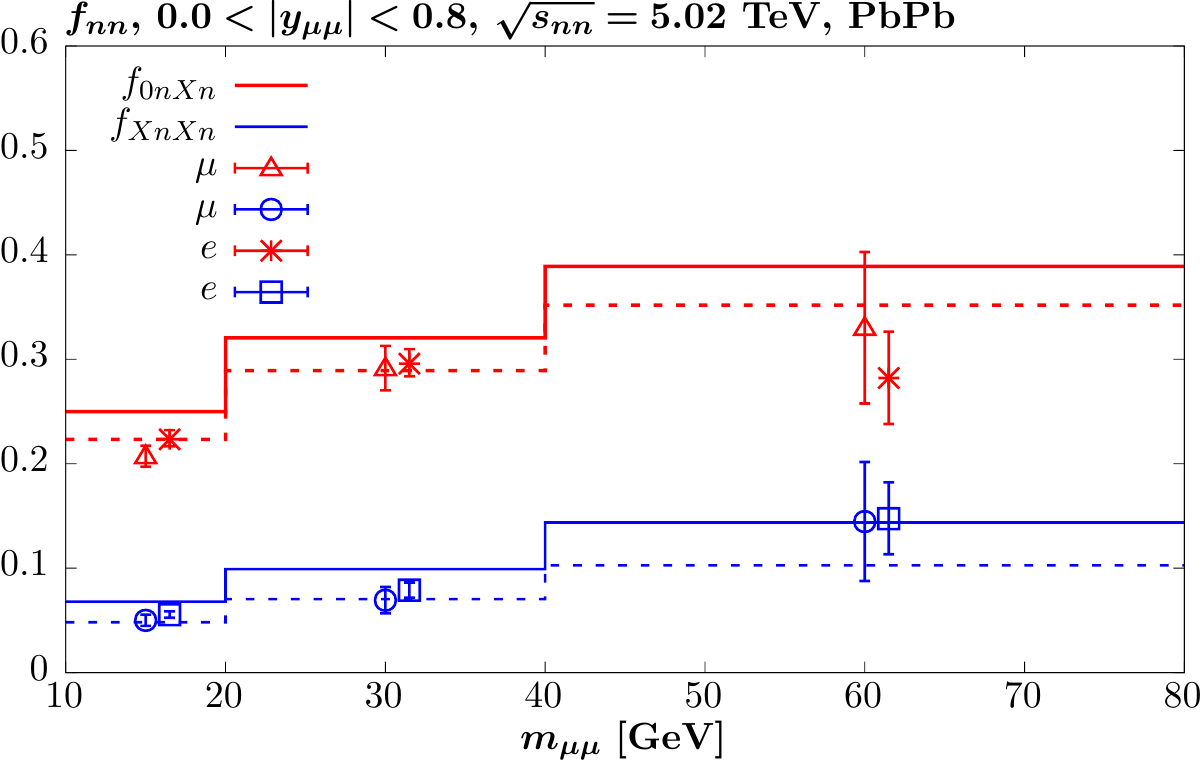}
\includegraphics[scale=0.6]{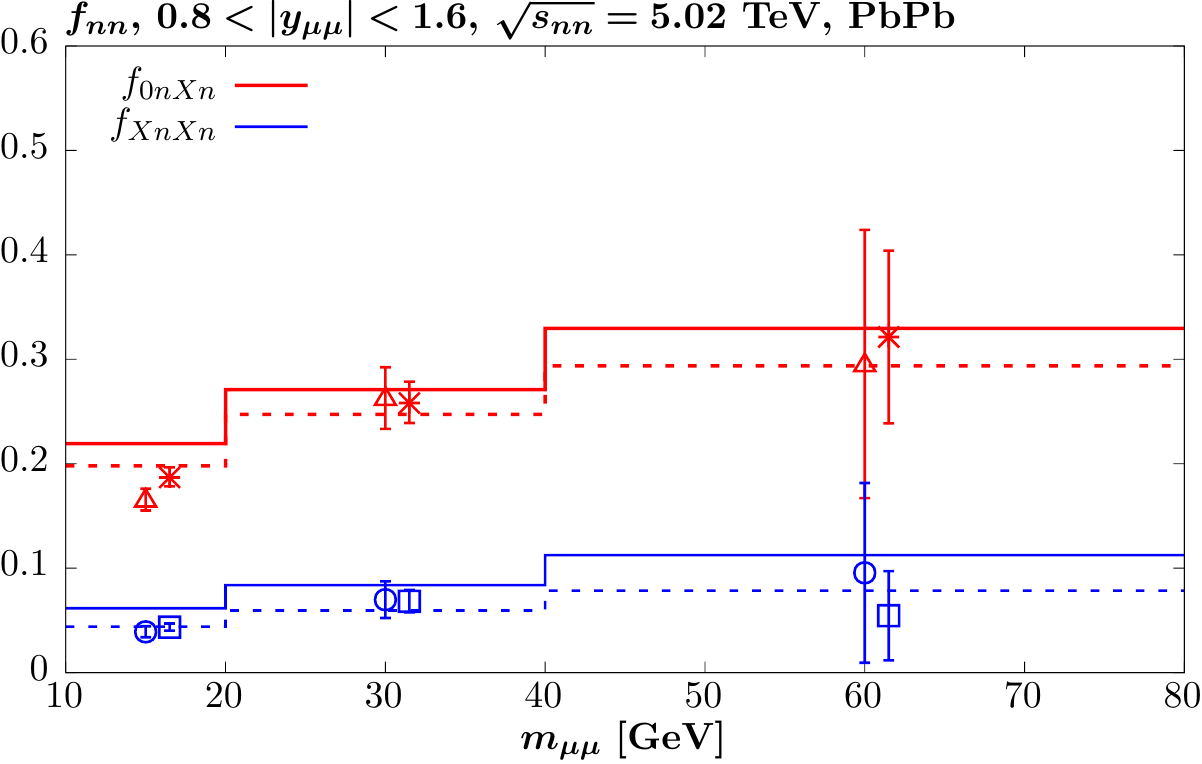}
\includegraphics[scale=0.6]{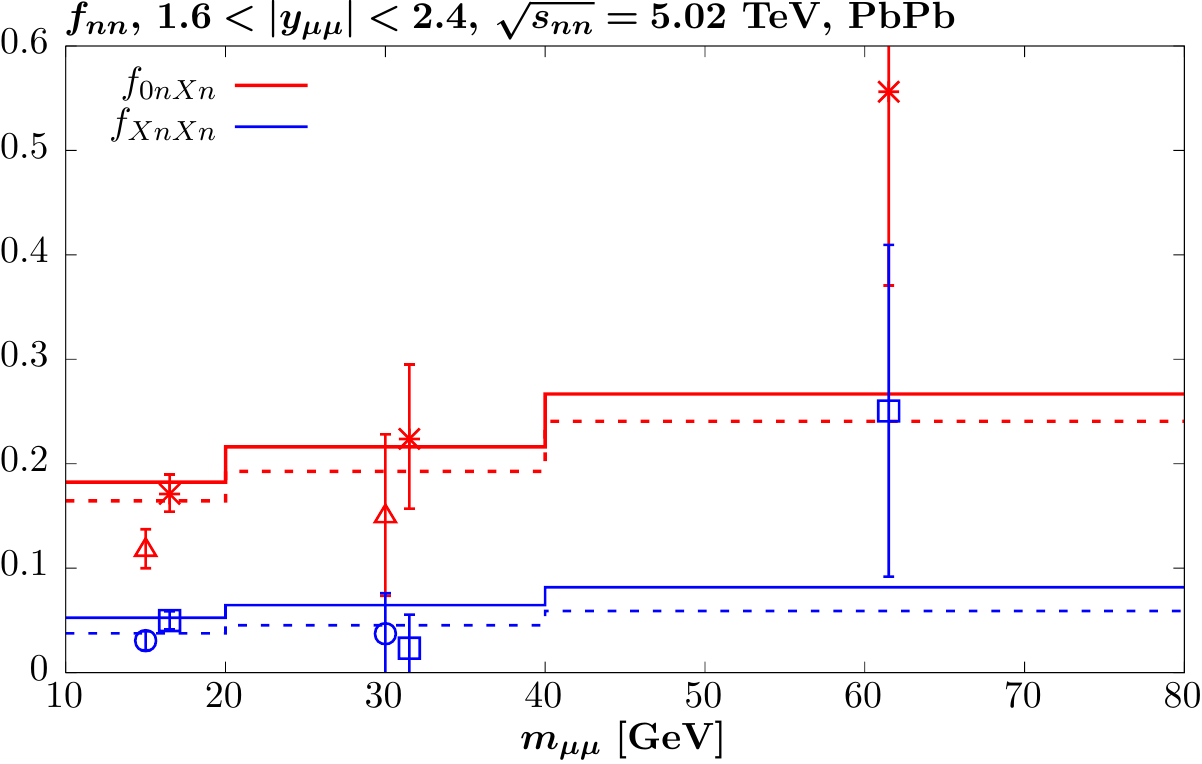}
\caption{As in Fig.~\ref{fig:ATLASrap} but with the ATLAS data in the dielectron channel~\cite{ATLAS:2022srr} for the same mass bins (and with a very similar event selection) shown. Theoretical predictions correspond to the dimuon event selection, but results for the dielectron case (which is very similar) are barely distinguishable, and hence are not shown for clarity. The solid histograms correspond to the default \texttt{SuperChic 4.2} predictions, while the dashed curves correspond to the result with the $\gamma A\to A^*$ cross section \eqref{eq:p1} multiplied by 0.8, for comparison.}
\label{fig:ATLAScomp}
\end{center}
\end{figure}

\begin{figure}
\begin{center}
\includegraphics[scale=0.6]{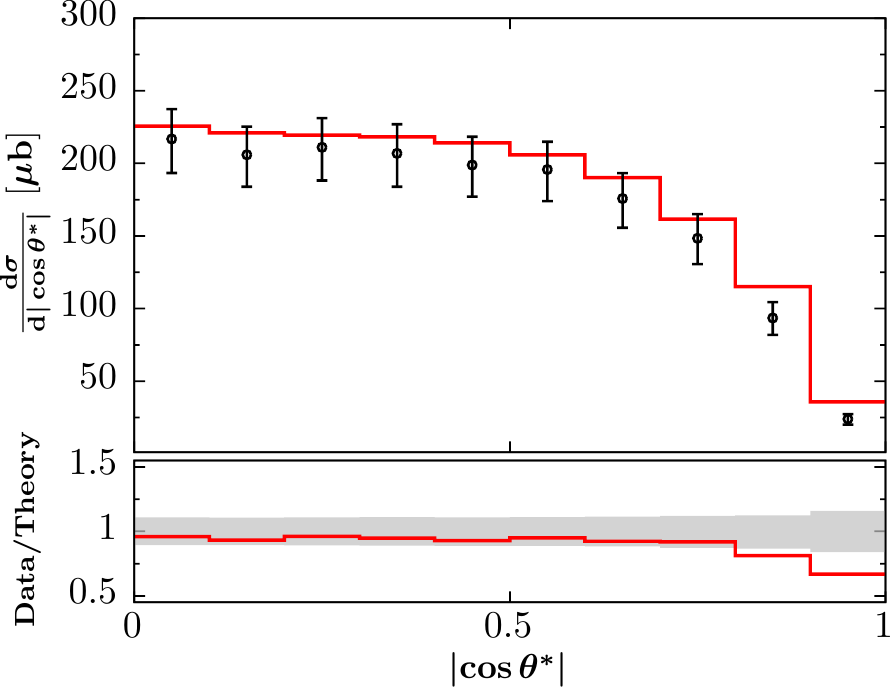}
\includegraphics[scale=0.6]{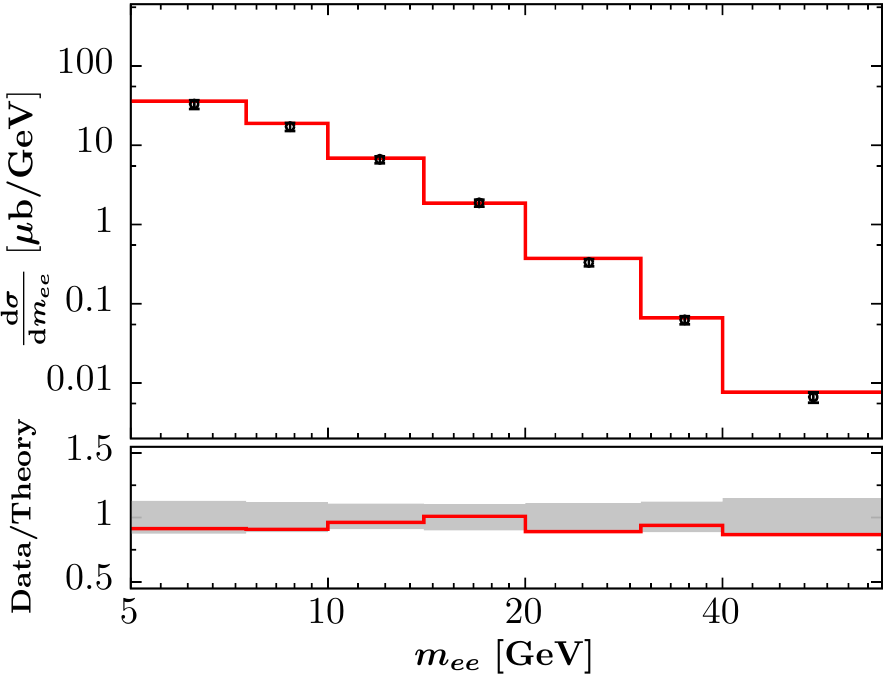}
\includegraphics[scale=0.6]{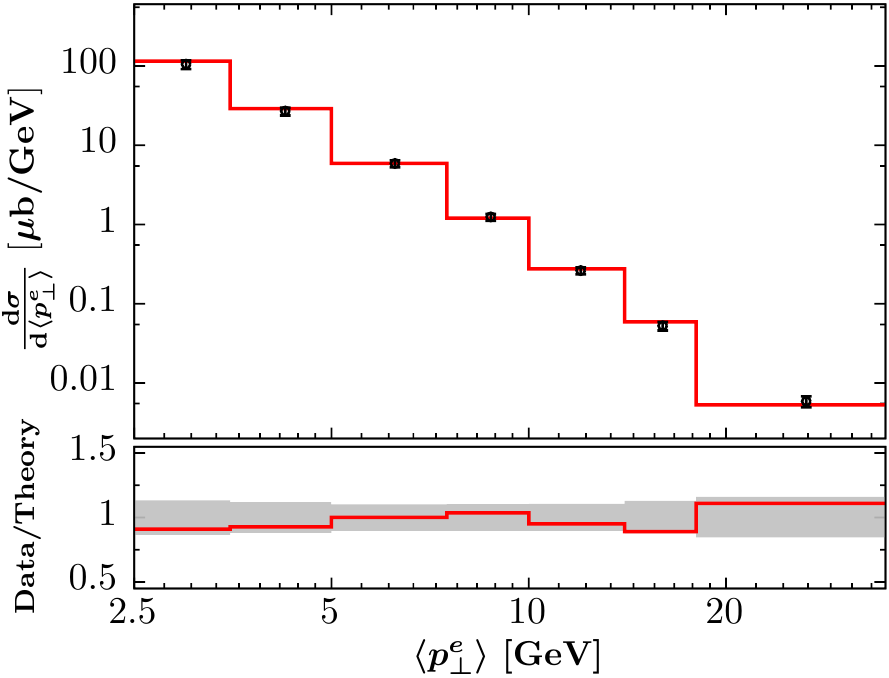}
\includegraphics[scale=0.6]{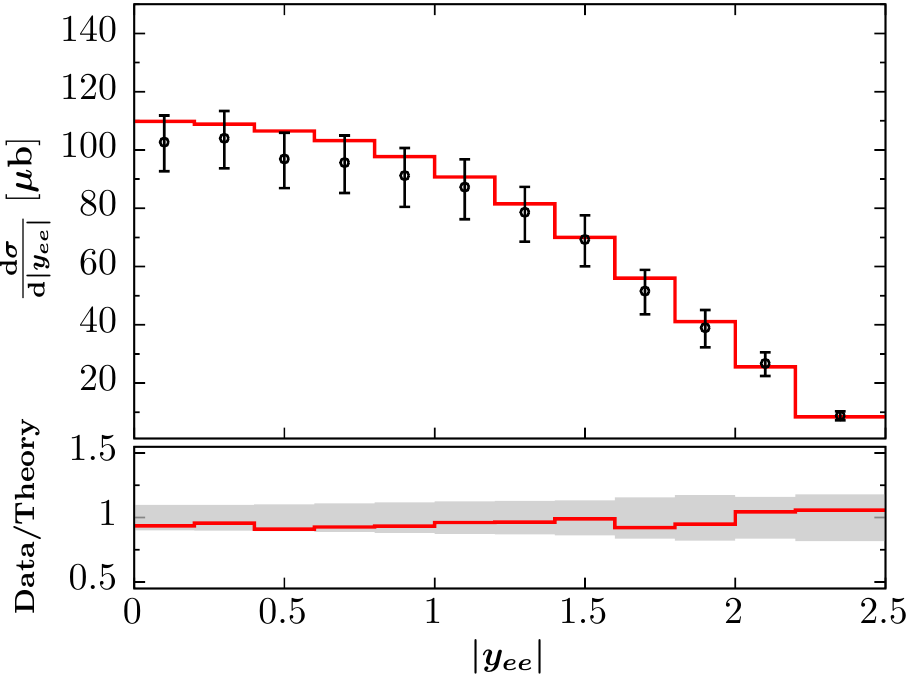}
\caption{Comparison of \texttt{SuperChic 4.2} predictions  to ATLAS data~\cite{ATLAS:2022srr} on ultraperipheral electron pair production in PbPb collisions at $\sqrt{s_{nn}}=5.02$ TeV in the $0n0n$ channel, and for a range of kinematic variables. The electrons are required to have $p_{\perp,e}>2.5$ GeV, $|\eta_e|<2.4$, $m_{ee} > 5$ GeV and $p_{\perp,\mu\mu}< 2$ GeV. Data errors correspond to systematic and statistical added in quadrature, and are shown by the grey band in the data/theory ratios.}
\label{fig:ATLASdieldist}
\end{center}
\end{figure}

Looking more closely, we can see that there is again a general trend to overshoot the $0nXn$, 	$XnXn$ data, and undershoot the $0n0n$ data, as there was in the dimuon data. Given, as discussed in Section~\ref{sec:iondiss}, there is a reasonable theoretical uncertainty in the predicted $\gamma A \to A^*$ cross section, it is interesting to investigate how much lower the input cross section would need to be in order to better match the data. Before doing so, we note that the second invariant mass bin, $10 < m_{ee} < 20$ GeV, in the dielectron measurement covers the same region as the lowest invariant mass bin in the dimuon measurement, see Figs.~\ref{fig:ATLASrap} and~\ref{fig:ATLASm}, and for the same dilepton rapidity region; the only difference from the point of view of the kinematic cuts is the tighter $p_\perp$ cut in the dimuon case. 
We would therefore expect rather similar fractions $f$ in both cases, and indeed that is true to very good approximation in the theoretical predictions. In terms of the data, on the other hand, the $0nXn$ and $XnXn$ fractions are rather higher in the dimuon case, at the  $\sim 2\sigma$ level. In Fig.~\ref{fig:ATLAScomp} we therefore show comparisons to both the dielectron and dimuon invariant mass distributions on the same plot. We can see that indeed the predicted $0nXn$, $XnXn$ distributions overshoot both sets of data, in particular at lower mass, but that this occurs rather more significantly for the dimuon data. We also show in the dashed histogram the predicted fractions when the default $\gamma A \to A^*$ cross section is multiplied by a factor of 0.8, and can see that in this case the agreement is rather better. It may therefore be that some amount of tuning is required in the future to better match the data. To enable this, in the public \texttt{SuperChic 4.2}  release we provide a flag (\texttt{fracsigX}) by which the normalization of the  $\gamma A \to A^*$ cross section may be modified. We note however, that in practice a reduction in the $\gamma A \to A^*$ cross section cannot simply be achieved by e.g. removing the higher energy and less well constrained region, where a Regge theory parameterisation must be used; even removing the entire cross section above $\omega > 20$ GeV (corresponding to $|y_{\gamma n}| \sim 6.5$) only reduces the cross section by a further $\sim 10\%$. This therefore corresponds to a fairly significant reduction.

We note that in principle another variable that will impact on the predicted dissociation fractions is the treatment of the survival factor. For example, if the suppression due to this is increased and/or pushed to lower impact parameter values, this will modify the average impact parameter sampled in the cross section. Given the dissociation probabilities have distinct impact parameter dependencies as in Fig.~\ref{fig:PX}, this will then modify these. However, on closer investigation we find that it is only with the rather extreme variations in the survival factor (of the type examined in~\cite{Harland-Lang:2021ysd}) that a noticeable reduction in the $0nXn$, $XnXn$ fractions occurs, and not necessarily with a better description of both cases simultaneously. A further way to shed light on this issue would be to present data in the $0n1n$ and $1n1n$ channels. In these cases the ion dissociation is guaranteed to be dominated by the GDR region, where uncertainties due to the high energy regime that enter the $Xn0n$ and $XnXn$ cross sections are absent. It would therefore be useful to determine whether the difference between data and theory persists in these single neutron channels, or is absent.

Finally, in Fig.~\ref{fig:ATLASdieldist} we compare predictions for a range of kinematic observables in the $0n0n$ case. Namely, the cosine of the electron scattering angle $\theta^*$ in the dielectron rest frame, the dielectron invariant mass and rapidity, and the average single electron/positron transverse momentum, $p_\perp^e$. We can see that in general the agreement between data and theory is very good, both at the level of normalization and shape; the $p_\perp^e$ and $m_{ee}$ distributions match well across roughly four orders of magnitude in the cross section. The largest discrepancy is in the $\cos \theta^*$ at forward angles, which may indicate some sensitivity to higher order QED effects in the $\gamma\gamma \to e^+ e^-$ process that are absent in the current theoretical treatment. In addition, we can see that in terms of the normalization there is a trend for the theoretical prediction to lie at the lower end of the data uncertainty region, though they are consistent within errors.

\subsection{Comparison to CMS data}\label{sec:CMS}

\begin{figure}
\begin{center}
\includegraphics[scale=0.62]{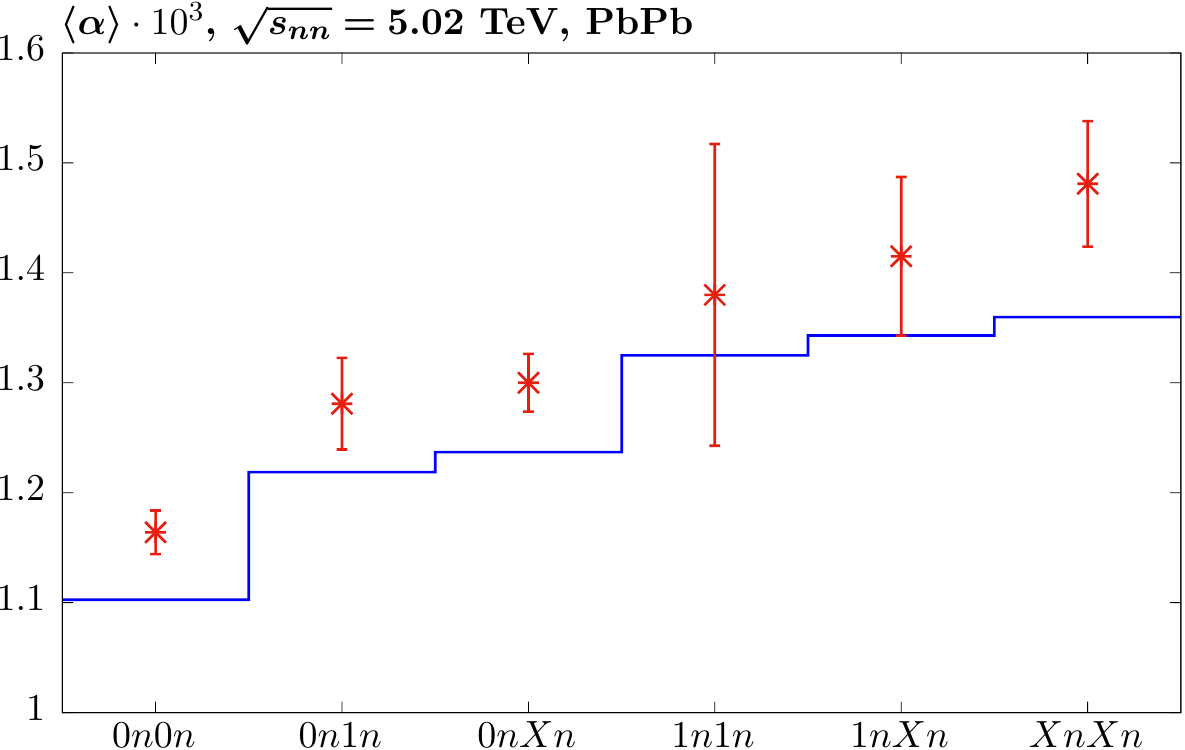}
\includegraphics[scale=0.62]{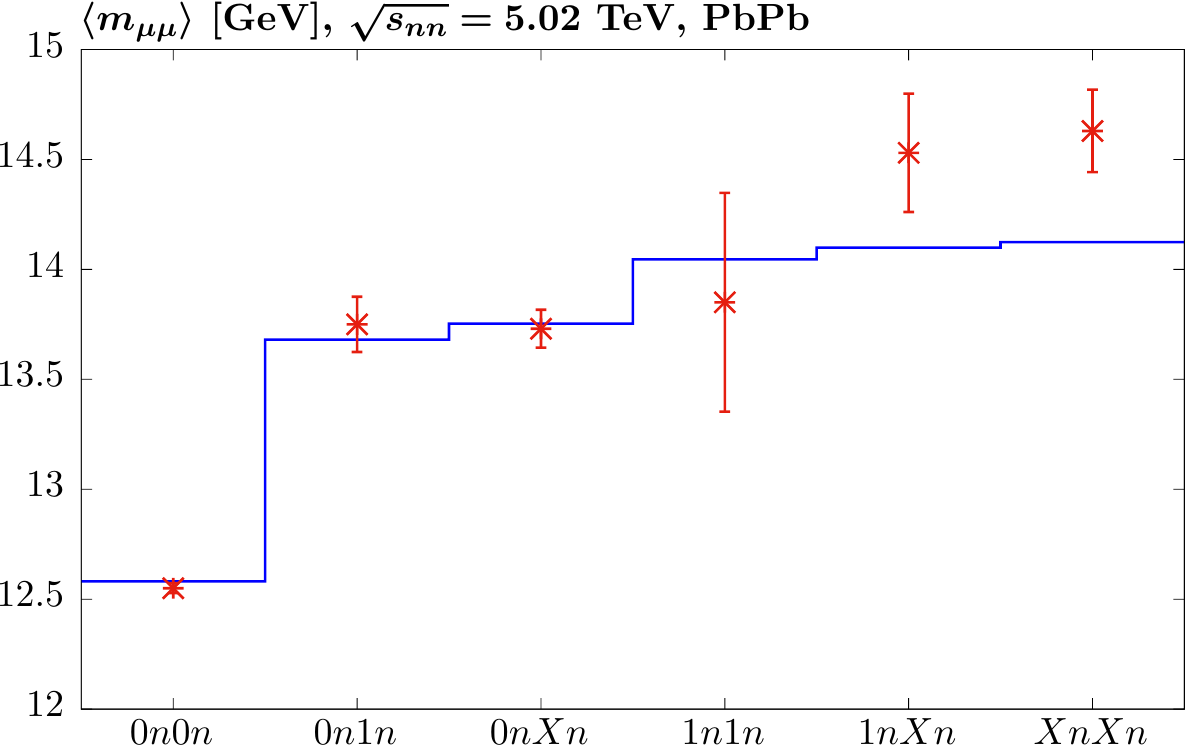}
\caption{Comparison of \texttt{SuperChic 4.2} predictions  to CMS data~\cite{CMS:2020skx} on ultraperipheral muon pair production in PbPb collisions at $\sqrt{s_{nn}}=5.02$ TeV for the average dimuon acoplanarity, $\alpha$, and invariant mass, $m_{ee}$, for different neutron tags. The muons are required to have $p_{\perp,\mu}>3.5$ GeV, $|\eta_\mu|<2.4$, $8< m_{\mu\mu} < 60$ GeV. Data errors correspond to systematic and statistical added in quadrature.}
\label{fig:CMSav}
\end{center}
\end{figure}

In this section we compare to CMS measurement~\cite{CMS:2020skx}  of muon pair production in ultraperipheral PbPb collisions at $\sqrt{s_{nn}}=5.02$ TeV. These are presented for a wide range of neutron tags, namely the $0n0n$, $0n1n$, $0nXn$, $1n1n$, $Xn1n$ and $XnXn$ final states. The acoplanarity, $\alpha = 1-\Delta\phi_{\mu\mu}/\pi$, distributions are measured over a wide range of $\alpha$ and fitted with a `core' and `tail' distribution, where the former is taken to be dominated by the leading order $\gamma\gamma \to \mu^+\mu^-$ process, and the latter is sensitive to higher order QED effects such as photon radiation from the dimuon system. The core distributions are then used to extract average acoplanarities for each neutron tag, as well as the average invariant dimuon invariant mass, comparisons to which as shown in Fig.~\ref{fig:CMSav}. 

In the Fig.~\ref{fig:CMSav} (left) the average acoplanarity is shown. We can see that the broad trend that is predicted is for this to increase as one requires more neutron emission, moving from the $0n0n$ to the $XnXn$ cases. This is exactly as expected from the discussion in the previous section, namely that as in Fig.~\ref{fig:PX} the $0n0n$ ($XnXn$) dissociation probability is peaked least (most) strongly at lower ion--ion impact parameters $b_\perp$. Given this we expect the, Fourier conjugate, dimuon transverse momentum $p_\perp^{\mu\mu}$ to be peaked most strongly towards lower (higher) values in the $0n0n$ ($XnXn$) case, with the intermediate cases lying between these extremes. This is precisely as observed in  Fig.~\ref{fig:CMSav} (left), bearing in mind that the dimuon acoplanarity increases directly with $p_\perp^{\mu\mu}$. This trend is also clearly seen in the data, validating this prediction. We note that this result only comes after including a full treatment of ion dissociation in transverse momentum space, as described in Section~\ref{sec:iondiss}. Indeed, in the CMS analysis~\cite{CMS:2020skx}  the results are compared to the \texttt{Starlight} MC~\cite{Klein:2016yzr}, where this is absent, and a rather flat scaling is predicted, in contradiction to the data. A similar level of agreement to the current case is on the other hand seen when comparing to the prediction of~\cite{Brandenburg:2020ozx}, which applies a similar framework to us, albeit without a full MC implementation. 

\begin{figure}
\begin{center}
\includegraphics[scale=0.62]{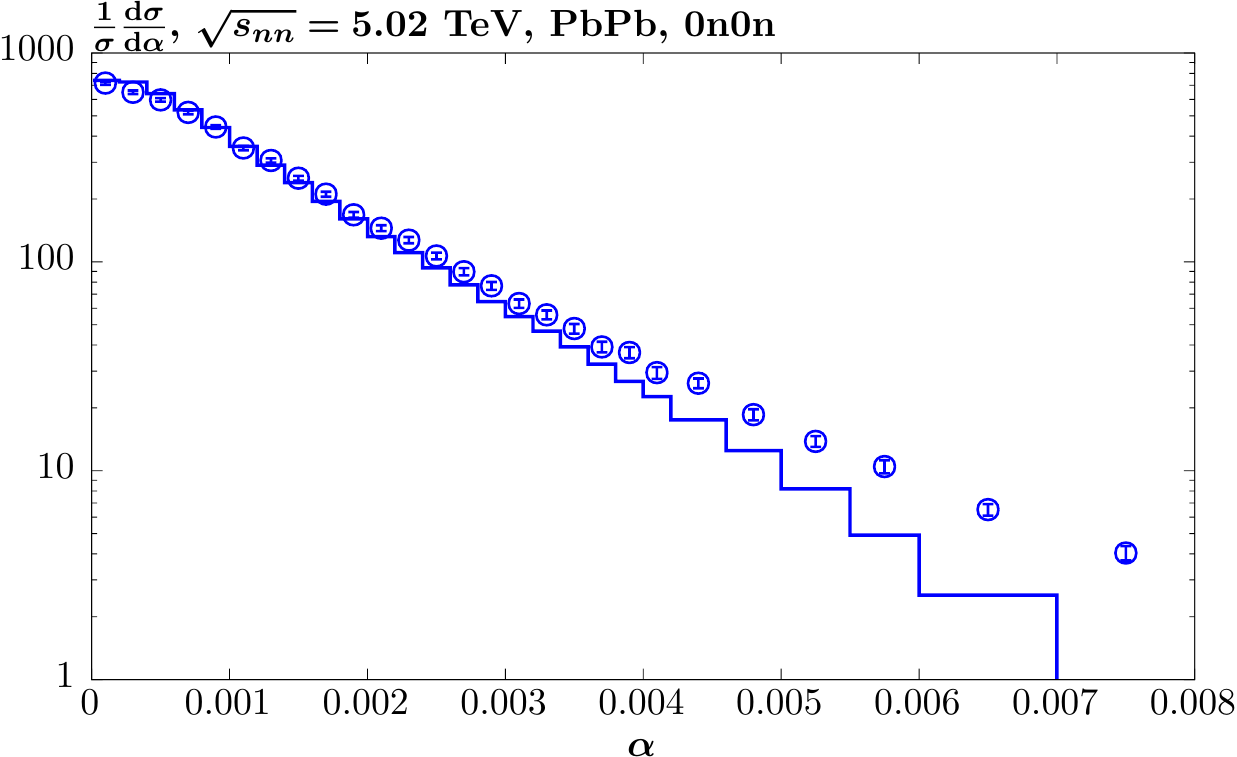}
\includegraphics[scale=0.62]{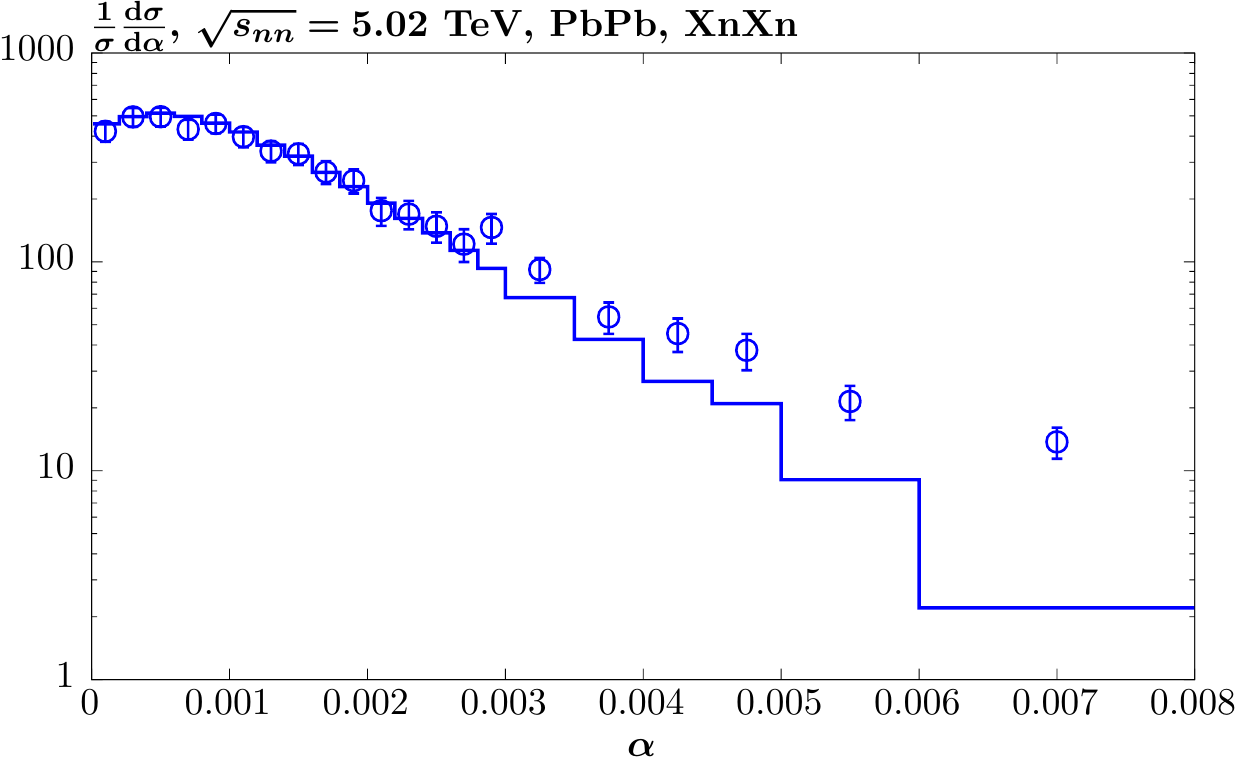}
\caption{Comparison of \texttt{SuperChic 4.2} predictions  to CMS data~\cite{CMS:2020skx} on ultraperipheral muon pair production in PbPb collisions at $\sqrt{s_{nn}}=5.02$ TeV as a function of the dimuon acoplanarity, $\alpha$, for different neutron tags. The distributions are defined such that the cross section is normalized in the $\alpha < 0.004$ region, where higher order QED effects are less significant. The muons are required to have $p_{\perp,\mu}>3.5$ GeV, $|\eta_\mu|<2.4$, $8< m_{\mu\mu} < 60$ GeV. Data errors correspond to systematic and statistical added in quadrature.}
\label{fig:CMSaco}
\end{center}
\end{figure}

While the overall trend in the data is consistent with our predictions, we can see that there is a systematic offset between data and theory, with the data having a somewhat larger average acoplanarity across all neutron tags. To understand this better, in Fig.~\ref{fig:CMSaco} we show the acoplanarity distributions for the two extreme, $0n0n$ and $XnXn$, cases. The data and theory predictions are defined such that the cross section is normalized in the $\alpha < 0.004$ region, where we expect higher order QED effects due to e.g. photon FSR to be less significant, see e.g. Fig.~12 of~\cite{ATLAS:2022srr}. We can that at lower acoplanarity the data are described rather well, but that beyond $\alpha \sim 0.002-0.004$ a clear excess in the data becomes apparent, which increases with increasing $\alpha$. It is therefore  possible that the average acoplanarity, as derived by the `core' distribution fits in the CMS analysis (which from Fig.~1 of~\cite{CMS:2020skx} describe the data well out to $\alpha \sim 0.004$), is being driven to somewhat larger values by such higher order QED effects. As these are absent in the current theoretical treatment, this would then lead the observed excess in Fig.~\ref{fig:CMSav} (left).

In Fig.~\ref{fig:CMSav} (right) the average dimuon invariant mass is shown. The basic trend is for this to increase as one requires more neutron emission, and is again  as expected from the discussion in the previous section, and indeed observed in the ATLAS data, for which the $0nXn$ and $XnXn$ event fractions are enhanced at larger dilepton invariant masses. This is therefore again an encouraging validation of the overall approach. Some excess of data over theory is on the other hand observed in the  $0nXn$ and $XnXn$ cases, albeit within relatively large experimental errors. 

\subsection{Comparison to STAR data}

\begin{figure}
\begin{center}
\includegraphics[scale=0.62]{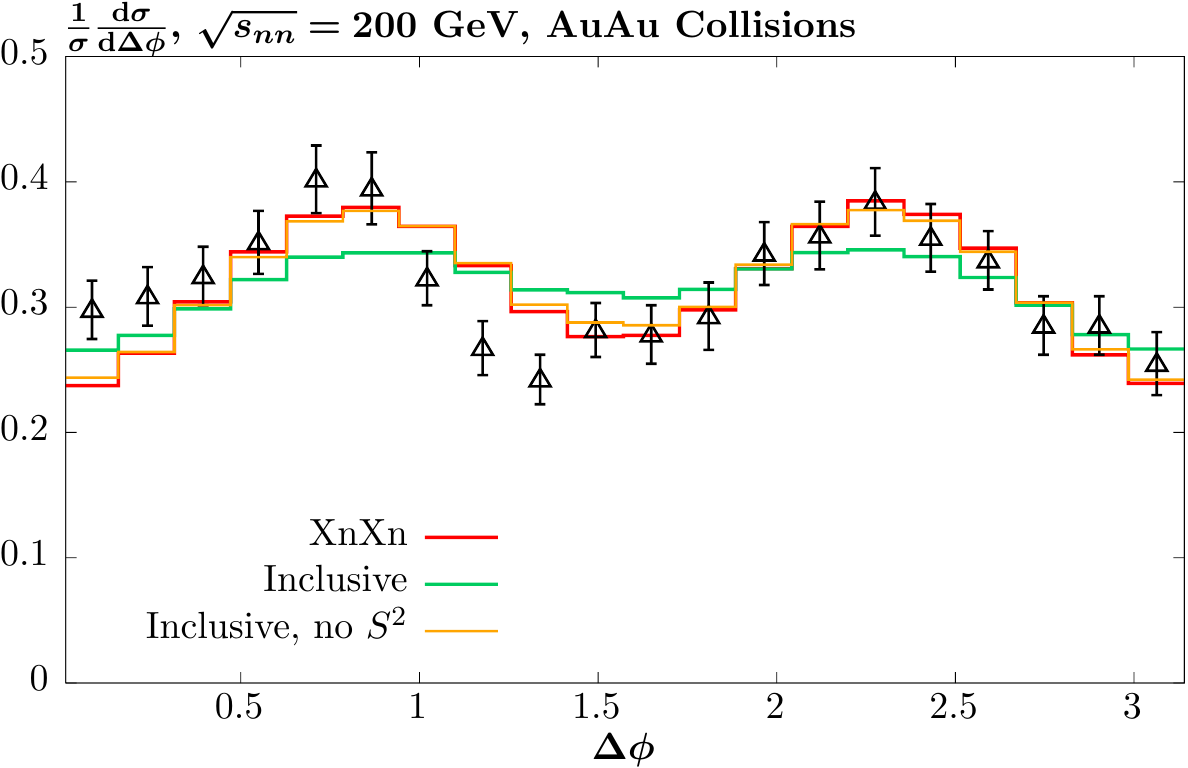}
\includegraphics[scale=0.62]{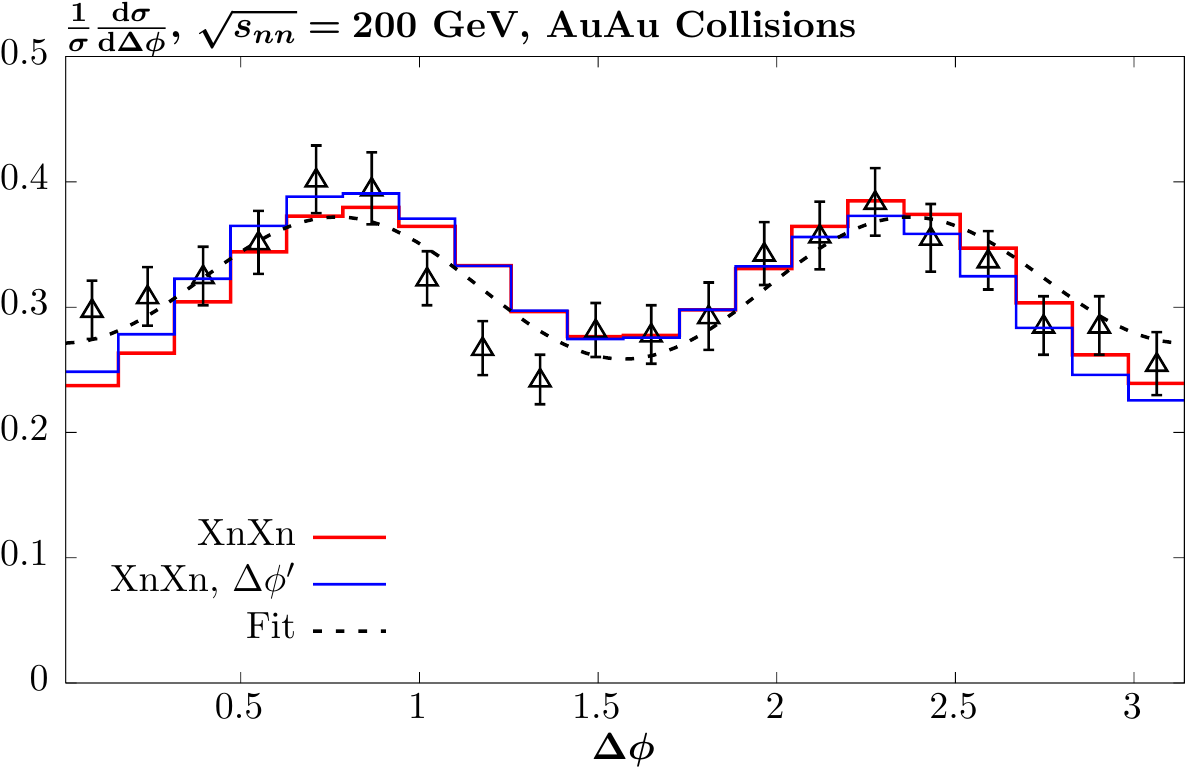}
\caption{Comparison of \texttt{SuperChic 4.2} predictions  to STAR data~\cite{STAR:2019wlg} on ultraperipheral electron pair production in AuAu collisions at $\sqrt{s_{nn}}=200$ GeV as function of the azimuthal angular separation, $\Delta\phi$, defined in the text. In the left plot the prediction for the inclusive (with respect to the neutron tag) case, both with and without including the ion--ion survival factor, are shown for comparison. In the right plot the predicted distribution with respect to the alternative variable, $\Delta\phi^\prime$, defined in the text, as well as the fit of the functional form \eqref{eq:dphistar} to the data are shown. The electrons are required to have $p_{\perp,e}>0.2$ GeV, $|\eta_e|<1.0$, $0.4< m_{ee} < 2.6$ GeV, and $p_{\perp,ee}< 0.1$ GeV. Data errors correspond to systematic and statistical added in quadrature.}
\label{fig:starphi}
\end{center}
\end{figure}

Finally, in this section we compare to the STAR measurement~\cite{STAR:2019wlg} of ultraperipheral electron pair production in AuAu collisions at $\sqrt{s_{nn}}=200$ GeV. These data are taken with a $XnXn$ neutron tag imposed, or more precisely a $YnYn$ tag with $Y=1,2,3$ suitably corrected up to the full $X > 0$ case. A particular observable of interest is the azimuthal angle $\Delta\phi$, defined in~\cite{STAR:2019wlg} as the angle  between the dielectron transverse momentum, $p_\perp^{ee}$, and the transverse momenta of one of the $e^\pm$. It is in particular predicted in~\cite{Li:2019yzy} (see also~\cite{Li:2019sin,Shao:2022stc,Xiao:2020ddm}) that  a modulation of the type
\be\label{eq:dphistar}
\frac{{\rm d}\sigma}{{\rm d}\Delta\phi} \propto 1 + A_{2\Delta\phi} \cos 2\Delta\phi+ A_{4\Delta\phi} \cos 4\Delta\phi\;,
\ee
should be observed in the case of dilepton UPCs, with $A_{2\Delta\phi}$ being zero up to lepton mass corrections $\sim m_l^2/m_{ll}^2$, and the precise value of $A_{4\Delta\phi}$ depending on the specific kinematics.
To be precise, the variable  $\Delta\phi$ is in fact defined in~\cite{Li:2019yzy} to be  the angle between the the dielectron transverse momentum and the vector difference ${\mathbf l_\perp} = \mathbf{p_\perp^{e^+}- p_\perp^{e^-}}$, which only coincides with the STAR experimental definition in the $p_{e_\perp} \gg p_\perp^{ee}$ limit. While this is true to reasonable approximation, it is as we will see not exact, and so we will for completeness consider both definitions, denoting that of~\cite{Li:2019yzy} as $ \Delta\phi$, and the STAR definition as $ \Delta\phi^\prime$.

In Fig.~\ref{fig:starphi} we compare our predicted normalized $\Delta \phi$ distribution to the STAR data, while in Table~\ref{tab:star} we compare the predicted values of $A_{(2,4)\Delta\phi}$, as well as the root mean squared dielectron transverse momentum, $p_\perp^{ee}$, and the fiducial cross section to the measured values. We first observe that the predicted total cross section is in excellent agreement within uncertainties with the data. Next, in Fig.~\ref{fig:starphi} (left) we also show for comparison the predicted distribution in the inclusive (with respect to neutron tag) case, with and without including the ion--ion survival factor. 
In general, we observe an oscillatory behaviour with respect to $\Delta \phi$, which arises due to the photon polarization dependence in the $\gamma\gamma\to X$ production amplitude (see~\eqref{eq:Agen}).
This effect was first discussed in~\cite{Khoze:2002dc} and is appropriately incorporated in \texttt{SuperChic}, but is not always accounted for in public MCs, such as e.g. \texttt{Starlight}~\cite{Baltz:2009jk}.

We can see by comparing to the inclusive case that the $XnXn$ tag requirement and the survival factor both have a non--negligible (and in fact counteracting) impact on the predicted distribution. This is  due to the differing impact parameter dependence of the ion dissociation probability with respect to the inclusive case, which modifies the corresponding amplitude \eqref{eq:tqts2} (after making the replacement \eqref{eq:gamrep}) in a non--trivial way. We can then see from \eqref{eq:Agen} that this amplitude couples the initial--state photon transverse momenta to the  $\gamma\gamma \to e^+ e^-$ vertex such that the weight of the contributing photon helicity amplitudes will be modified by the ion dissociation probability (as well as the ion--ion survival factor) and its particular impact parameter dependence. Indeed, a discussion of this effect from a different perspective is presented in~\cite{Li:2019yzy}. As an aside, we note that for the $1n1n$ case the predicted distribution is very similar (although not identical) to the $XnXn$ case, hence any potential effect from the fact that the ZDC tag does not extend beyond 3 neutrons should be negligible.

\begin{table}
\begin{center}
\begin{tabular}{|c|c|c|}
\hline
& \texttt{SuperChic}& Data \\
\hline
 $\sigma$ [$\mu$b] & 240 &    $261 \pm 37$ \\    
 \hline
   $| A_{2\Delta\phi}|$ (\%)&    6.2    &   $2.0\pm 2.4$     \\    
   $| A_{4\Delta\phi}|$ (\%)&    20.1 &      $ 16.8\pm 2.5$   \\    
    $\sqrt{\left\langle (p_\perp^{ee})^2\right \rangle}$ [MeV]&   36.1    &   $38.1\pm 0.9$     \\  
   \hline  
\end{tabular}
\caption{Predicted values of $A_{2,4\Delta\phi}$, as well as the root mean squared dielectron transverse momentum, $p_\perp^{ee}$, and the fiducial cross section to the STAR data~\cite{STAR:2019wlg}. The corresponding coefficients with respect to $\Delta \phi^\prime$ are very similar, and are therefore not given. Uncertainties correspond to sum in quadrature of all quoted sources from the experimental analysis.} \label{tab:star}
\end{center}
\end{table}

For the appropriate $XnXn$ case we can see in Fig.~\ref{fig:starphi} that the predicted distribution matches the data reasonably well. The agreement is in particular excellent in the $\Delta \phi \gtrsim \pi/2$ region, while below this there are some discrepancies. In the right plot, however we also show the result of a direct fit using \eqref{eq:dphistar} to the data, and while this achieves a somewhat better agreement at low $\Delta \phi$, overall the agreement is not significantly improved. We also show a comparison to the experimental definition, $ \Delta\phi^\prime$; we can that the predicted distribution is indeed mildly different, with a somewhat better agreement at lower $\Delta \phi$, but again overall the agreement is not significantly improved. 

Therefore, there is possibly a limit to how well any prediction can match the measurement, perhaps due to fluctuations in the data or some other systematic effect. Nonetheless, in Table~\ref{tab:star} the predicted values of $A_{2,4\Delta\phi}$ are compared to the values determined in~\cite{STAR:2019wlg} from a fit to the data and there is fair agreement, at the $\sim 1-2\sigma$ level.
With further more precise data the agreement may of course improve, but a more precise analysis, accounting for e.g. higher order QED effects may improve this. Indeed, in e.g.~\cite{Li:2019yzy,Klein:2020jom} the impact of accounting for photon FSR from the dilepton system is discussed and found to be non--negligible in some cases. Similarly, in Table~\ref{tab:star} the root mean squared dielectron transverse momentum is also given, and again found to be in fair agreement with the data, but to lie below the measured value; photon FSR effects will act precisely to increase this. 

\begin{figure}
\begin{center}
\includegraphics[scale=0.62]{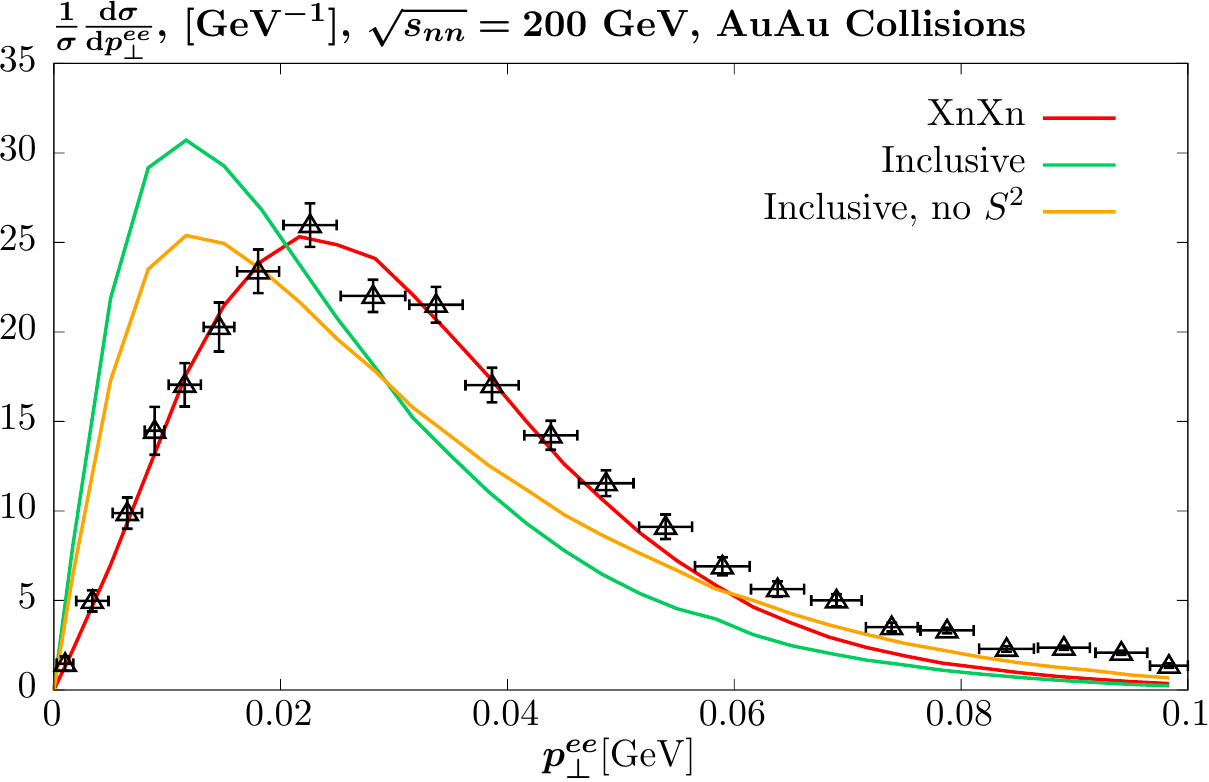}
\caption{Comparison of \texttt{SuperChic 4.2} predictions  to STAR data~\cite{STAR:2019wlg} on ultraperipheral electron pair production in AuAu collisions at $\sqrt{s_{nn}}=200$ GeV as function of the dielectron transverse momentum, $p_{\perp,ee}$. The electrons are required to have $p_{\perp,e}>0.2$ GeV, $|\eta_e|<1.0$, $0.4< m_{ee} < 2.6$ GeV, and $p_{\perp,ee}< 0.1$ GeV. Data points are extracted from~\cite{STAR:2019wlg}; as precise $p_\perp$ binning not publicly available theoretical results are presented as curves only.}
\label{fig:starpt}
\end{center}
\end{figure}

Indeed, in Fig.~\ref{fig:starpt} we compare the predicted transverse momentum distribution with the STAR data and the agreement is seen to be very good for most of the $p_{\perp,ee}$ region, but with some excess of data over theory at the larger values, which again is as we would expect from photon FSR effects. We also show for comparison the predicted distribution in the inclusive (with respect to neutron tag) case, with and without including the ion--ion survival factor. The impact of a full kinematic account of ion dissociation is again clear, and it is only after doing this that the data are matched well. The fact that the $XnXn$ tag leads to a broadening of the transverse momentum distribution towards higher values is again exactly as expected from the impact parameter dependence of the $XnXn$ dissociation probability, which is more strongly peaked to lower values, i.e. larger $p_{\perp,ee}$.

As mentioned above, in~\cite{Li:2019yzy,Xiao:2020ddm,Shao:2022stc} it is argued that the $A_{2\Delta\phi}$ coefficient in \eqref{eq:dphistar} should be zero up to $\sim m_e^2/m_{ee}^2$ electron mass corrections. However, we can see from Table~\ref{tab:star} that this is not the case for our prediction. The reason for this comes from considering precisely what is assumed in these analyses, namely that the photon transverse momenta are much smaller than the electron transverse momenta, i.e. $p_\perp^{ee}\ll p_\perp^e$. However, the STAR data extend down to  a minimum $p_\perp^e > 200$ MeV, which is indeed larger than the root mean squared dielectron transverse momentum, $\sqrt{\langle (p_\perp^{ee})^2\rangle } \sim 40$ MeV, but not so large that one can necessarily neglect the dielectron transverse momentum in the calculation. More precisely, the predicted distribution \eqref{eq:dphistar} in fact corresponds~\cite{Shao:2022stc} to the differential cross section with respect to  ${\mathbf p_\perp^{ee}}$ and the vector difference ${\mathbf l_\perp}$ defined above, whereas the observed cross section is of course integrated over these. The lepton transverse momenta cuts $p_\perp^e > p_\perp^{\rm cut} = 200$ MeV correspond in terms of these to
\be\label{eq:pcut}
l_\perp^2 + (p_\perp^{ee})^2 \pm 2 |l_\perp||p_\perp^{ee}| \cos \Delta\phi > 4(p_\perp^{\rm cut})^2 \;,
\ee
which can introduce a dependence on $\cos\Delta\phi$, that is not captured by \eqref{eq:dphistar} with $A_{2\Delta\phi}=0$. The precise form of this depends on the above cut and its non--trivial interplay with the full kinematic dependence of the production cross section, and hence is not straightforward to predict analytically. It is only by providing a full MC treatment, as we do here, that this can be accounted for. We note that if we remove the $p_\perp^e$ cuts, then the predicted value of $A_{2\Delta\phi}$ is indeed consistent with zero, as it is if we increase the threshold on $m_{ee}$ to e.g $\sim 4$ GeV; this effect is therefore rather specific to the STAR kinematics. If we reduce the electron mass arbitrarily, then this makes a negligible difference, confirming that this is not the relevant factor.

In the STAR analysis~\cite{STAR:2019wlg} the data are compared to a `QED' prediction from~\cite{Zha:2018tlq}. Although the language used is sometimes different, the basic approach of this is the same as that applied here, i.e. the impact parameter dependent ion dissociation and ion--ion survival probabilities are accounted for and appropriately translated to transverse momentum space, with the standard LO QED $\gamma\gamma \to l^+ l^-$ amplitude applied and the ion photon flux accounted for via the usual ion EM form factors. Qualitatively we see reasonable agreement with these results, but they are not identical. The reason for this is unclear, and may lie in the precise implementation of the above theoretical ingredients. However, we note that the quoted value of $A_{2\Delta\phi}$ is indeed exactly zero, contrary to the discussion above, and therefore these predictions must rely on the $p_\perp^{ee}\ll p_\perp^e$ approximation, which as discussed above is not necessarily valid for the STAR data. Hence, this is a possible reason for the observed difference.

Finally, we end this section by noting that some care is needed in interpreting the comparisons made in the STAR analysis~\cite{STAR:2019wlg}. In particular, in the  previous versions of \texttt{SuperChic}, as is clearly described in~\cite{Harland-Lang:2018iur} for the case of version 3 where heavy ion UPCs are first considered, ion dissociation had not yet been implemented. That is, only inclusive (with respect to ion dissociation) production could be generated. The STAR data are on the other hand  taken with a $XnXn$ tag, that is they are not inclusive with respect to ion dissociation. Given this, and as we have discussed in detail above, we would  expect the \texttt{SuperChic} 3  predictions not to match the shape of the data distributions; this is precisely what is observed in Fig.~\ref{fig:starphi} (left) and Fig.~\ref{fig:starpt}. However, in~\cite{STAR:2019wlg}, comparisons are presented to  \texttt{SuperChic} 3, and this difference between what the data correspond to and what is actually generated by the Monte Carlo is not highlighted in the comparisons\footnote{In more detail, it would appear from the results presented in~\cite{STAR:2019wlg} that these correspond to the \texttt{SuperChic} predictions without the ion--ion survival factor included, i.e. not to a UPC.}. Indeed, the observed discrepancy between the \texttt{SuperChic} 3 predictions and the data is claimed in~\cite{STAR:2019wlg} to invalidate  the theoretical approach described in~\cite{Harland-Lang:2018iur}.  
It should be clear from the discussion here that this is not the case. It is furthermore stated that the treatment of~\cite{Harland-Lang:2018iur} neglects the fact that  energy spectrum of the colliding photons depends on the nucleus--nucleus impact parameter, and therefore, on the spatial distribution of the electromagnetic fields, with the implication being that this is the reason for the discrepancy. This is not correct: as discussed in detail in Section~\ref{sec:theory}, this dependence (i.e. the correlation between the ion--ion impact parameter and the measured kinematic distributions) is fully accounted for. It is simply that mutual ion--ion dissociation was not yet accounted for in  \texttt{SuperChic} 3, and hence it should not have been used for comparison to the STAR data. 

The above  confusion  appears to motivate the discussion in~\cite{Brandenburg:2021lnj}, where the approach presented here and in previous studies is incorrectly associated with the equivalent photon approximation (as discussed in Section~\ref{sec:key} this approximation is not applied in our calculation) and an apparently artificial distinction made between the approach discussed here and the result of a `QED' calculation~\cite{Zha:2018tlq,Brandenburg:2020ozx}. This appears to be based on the incorrect assumption that the method outlined here cannot account for impact parameter constraints of the sort implied by mutual ion--dissociation conditions (see in particular the discussion in Section 3.3 of~\cite{Brandenburg:2021lnj}). It should be clear from the results presented in this paper that this is not the case. The `QED' results of~\cite{Zha:2018tlq,Brandenburg:2020ozx} and those of this work are built from precisely the same underlying ingredients, and account for  the correlation between the ion--ion impact parameter and the measured kinematic distributions in the same manner. Indeed, the predictions of~\cite{Zha:2018tlq} with respect to both the STAR data and the CMS measurement presented in Section~\ref{sec:CMS} agree rather well with our results.

 \section{Summary and Outlook}\label{sec:conc}

Ultraperipheral heavy ion collisions (UPCs) provide a  source of photon--photon collisions with which we can probe new phenomena within and potentially beyond the Standard Model. However, to leverage this initial--state to its maximum potential requires that the underlying process be modelled with as much precision as possible. In this paper we have discussed a new development of an approach which aims to provide this. We have in particular presented an update to the \texttt{SuperChic} MC generator to account for mutual ion dissociation in UPCs, due to additional photon exchanges between the colliding ions, whereby one or both ions can be excited into a higher energy state that subsequently decays by emitting a single or multiple neutrons. This is released in the \texttt{SuperChic 4.2} MC, the code and a user manual for which can be found at 
\\
\\
\href{https://superchic.hepforge.org}{\tt http://projects.hepforge.org/superchic}
\\
\\
This simulates a range of exclusive processes in both heavy ion and proton collisions, see~\cite{Harland-Lang:2018iur,Harland-Lang:2020veo} for more details. 

The theoretical framework presented in this paper can provide a full account of the relevant physics effects in UPCs. In particular, while the survival factor probability of no additional inelastic ion--ion scattering due to the strong interaction, and the mutual ion dissociation probability, are both most straightforwardly formulated in the impact parameter space of the colliding ions, we have discussed how these can consistently be accounted for and translated to momentum space. In this way we can systematically account for the modifications these entail for the measured kinematic distributions of the centrally produced state. While the importance of such a systematic treatment in the case of the survival factor has been discussed elsewhere (see e.g.~\cite{Harland-Lang:2021ysd}), here we have focused on the case of mutual ion dissociation, which is included for the first time in \texttt{Superchic}.

This is of particular relevance given it is possible to measure such ion dissociation processes via zero degree calorimeter (ZDC) detectors at the LHC and RHIC. The previous version 3 of \texttt{Superchic} is inclusive with respect to ion dissociation, and hence can be reliably used when the data are also presented inclusively. However, it is possible with ZDCs to presented data for a range of neutron tags, that is with and without ion dissociation requirements in place. Such processes can now be modelled consistently within the same overall framework. 

We have compared to a range of data from ATLAS, CMS and STAR and demonstrated how these ion dissociation requirements (or vetos) in the ZDCs have a non--negligible impact on the predicted kinematic distributions, and observed rather good agreement between the data and theory for electron and muon pair production. 
On the other hand, some apparent discrepancies exist in certain regions of phase space. Some of these we can understand to be due to higher order QED corrections, in particular due to photon FSR from the leptons. A full account of this, and indeed other higher QED effects due to lepton--ion interactions and unitary corrections relating to additional electron pair production, is therefore well motivated as future work, and is left to that. Further mild differences may indicate the need for further tuning in the input $\gamma A \to A^*$ cross sections, and we provide an approximate method to achieve this directly in the \texttt{SuperChic 4.2} code. Further data on e.g. the $1n0n$ and $1n1n$ channels, as well as the $Xn0n$, $XnXn$ channels will shed more light on this.

In summary, in this work we have presented a significant update to the \texttt{Superchic} MC to account for mutual ion dissociation, allowing this to be used for the first time to compare against the range of data that can be (and has been) collected in high energy heavy ion collisions with ZDC tags in place. The agreement between data and theory for electron and muon pair production is observed to be in general good, providing key support for the approach presented here and motivation to use it to interpret data on less well tested final states, such as $\tau$ leptons (and the implications for the anomalous magnetic moment of the $\tau$) and light--by--light scattering.

\section*{Acknowledgements}

I am grateful to Valery Khoze and Misha Ryskin for comments on the manuscript and  useful discussions. I am grateful to Shi Pu and the authors of~\cite{Wang:2022gkd} for providing the STAR~\cite{STAR:2019wlg}  datapoints for $\Delta \phi$ distribution, and  Iwona Grabowska--Bold for providing data from the ATLAS dielectron analysis~\cite{ATLAS:2022srr} as well as useful discussions. I thank  the Science and Technology Facilities Council (STFC) part of U.K. Research and Innovation for support via the grant award ST/T000856/1.

\bibliography{references}{}

\begin{thebibliography}{10}

\bibitem{Braun-Munzinger:2015hba}
P.~Braun-Munzinger, V.~Koch, T.~Sch\"afer, and J.~Stachel,
\newblock Phys. Rept. {\bf 621}, 76 (2016), 1510.00442.

\bibitem{Busza:2018rrf}
W.~Busza, K.~Rajagopal, and W.~van~der Schee,
\newblock Ann. Rev. Nucl. Part. Sci. {\bf 68}, 339 (2018), 1802.04801.

\bibitem{ALICE:2022wpn}
ALICE,
\newblock (2022), 2211.04384.

\bibitem{Bruce:2018yzs}
R.~Bruce {\em et~al.},
\newblock J. Phys. G {\bf 47}, 060501 (2020), 1812.07688.

\bibitem{ATLAS:2017fur}
ATLAS, M.~Aaboud {\em et~al.},
\newblock Nature Phys. {\bf 13}, 852 (2017), 1702.01625.

\bibitem{CMS:2018erd}
CMS, A.~M. Sirunyan {\em et~al.},
\newblock Phys. Lett. B {\bf 797}, 134826 (2019), 1810.04602.

\bibitem{ATLAS:2019azn}
ATLAS, G.~Aad {\em et~al.},
\newblock Phys. Rev. Lett. {\bf 123}, 052001 (2019), 1904.03536.

\bibitem{ATLAS:2020hii}
ATLAS, G.~Aad {\em et~al.},
\newblock JHEP {\bf 11}, 050 (2021), 2008.05355.

\bibitem{Harland-Lang:2022end}
L.~A. Harland-Lang and M.~Tasevsky,
\newblock (2022), 2208.10526.

\bibitem{Biloshytskyi:2022dmo}
V.~Biloshytskyi {\em et~al.},
\newblock Phys. Rev. D {\bf 106}, L111902 (2022), 2207.13623.

\bibitem{Beresford:2019gww}
L.~Beresford and J.~Liu,
\newblock Phys. Rev. D {\bf 102}, 113008 (2020), 1908.05180.

\bibitem{Dyndal:2020yen}
M.~Dyndal, M.~Klusek-Gawenda, M.~Schott, and A.~Szczurek,
\newblock Phys. Lett. B {\bf 809}, 135682 (2020), 2002.05503.

\bibitem{ATLAS:2022ryk}
ATLAS,
\newblock (2022), 2204.13478.

\bibitem{CMS:2022arf}
CMS,
\newblock (2022), 2206.05192.

\bibitem{Bertulani:1987tz}
C.~A. Bertulani and G.~Baur,
\newblock Phys. Rept. {\bf 163}, 299 (1988).

\bibitem{Klein:2020fmr}
S.~Klein and P.~Steinberg,
\newblock Ann. Rev. Nucl. Part. Sci. {\bf 70}, 323 (2020), 2005.01872.

\bibitem{Klein:2016yzr}
S.~R. Klein, J.~Nystrand, J.~Seger, Y.~Gorbunov, and J.~Butterworth,
\newblock Comput. Phys. Commun. {\bf 212}, 258 (2017), 1607.03838.

\bibitem{Harland-Lang:2018iur}
L.~A. Harland-Lang, V.~A. Khoze, and M.~G. Ryskin,
\newblock Eur. Phys. J. {\bf C79}, 39 (2019), 1810.06567.

\bibitem{Burmasov:2021phy}
N.~Burmasov, E.~Kryshen, P.~Buehler, and R.~Lavicka,
\newblock Comput. Phys. Commun. {\bf 277}, 108388 (2022), 2111.11383.

\bibitem{Shao:2022cly}
H.-S. Shao and D.~d'Enterria,
\newblock JHEP {\bf 09}, 248 (2022), 2207.03012.

\bibitem{Harland-Lang:2021ysd}
L.~A. Harland-Lang, V.~A. Khoze, and M.~G. Ryskin,
\newblock SciPost Phys. {\bf 11}, 064 (2021), 2104.13392.

\bibitem{SuperCHIC}
The SuperCHIC code and documentation are available at {\tt
  http://projects.hepforge.org/superchic/}.

\bibitem{ALICE:2013wjo}
ALICE, E.~Abbas {\em et~al.},
\newblock Eur. Phys. J. C {\bf 73}, 2617 (2013), 1305.1467.

\bibitem{ATLAS:2020epq}
ATLAS, G.~Aad {\em et~al.},
\newblock Phys. Rev. C {\bf 104}, 024906 (2021), 2011.12211.

\bibitem{ATLAS:2022srr}
ATLAS,
\newblock (2022), 2207.12781.

\bibitem{CMS:2020skx}
CMS, A.~M. Sirunyan {\em et~al.},
\newblock Phys. Rev. Lett. {\bf 127}, 122001 (2021), 2011.05239.

\bibitem{STAR:2019wlg}
STAR, J.~Adam {\em et~al.},
\newblock Phys. Rev. Lett. {\bf 127}, 052302 (2021), 1910.12400.

\bibitem{Berman:1975tt}
B.~L. Berman and S.~C. Fultz,
\newblock Rev. Mod. Phys. {\bf 47}, 713 (1975).

\bibitem{Broz:2019kpl}
M.~Broz, J.~G. Contreras, and J.~D. Tapia~Takaki,
\newblock Comput. Phys. Commun. {\bf 253}, 107181 (2020), 1908.08263.

\bibitem{Bertulani:2005ru}
C.~A. Bertulani, S.~R. Klein, and J.~Nystrand,
\newblock Ann. Rev. Nucl. Part. Sci. {\bf 55}, 271 (2005), nucl-ex/0502005.

\bibitem{Baltz:2007kq}
A.~J. Baltz,
\newblock Phys. Rept. {\bf 458}, 1 (2008), 0706.3356.

\bibitem{Baltz:2009jk}
A.~J. Baltz, Y.~Gorbunov, S.~R. Klein, and J.~Nystrand,
\newblock Phys. Rev. C {\bf 80}, 044902 (2009), 0907.1214.

\bibitem{Brandenburg:2020ozx}
J.~D. Brandenburg {\em et~al.},
\newblock (2020), 2006.07365.

\bibitem{Li:2019yzy}
C.~Li, J.~Zhou, and Y.-J. Zhou,
\newblock Phys. Lett. B {\bf 795}, 576 (2019), 1903.10084.

\bibitem{Klein:2020jom}
S.~Klein, A.~H. Mueller, B.-W. Xiao, and F.~Yuan,
\newblock Phys. Rev. D {\bf 102}, 094013 (2020), 2003.02947.

\bibitem{Wang:2021kxm}
R.-j. Wang, S.~Pu, and Q.~Wang,
\newblock Phys. Rev. D {\bf 104}, 056011 (2021), 2106.05462.

\bibitem{Mazurek:2021ahz}
K.~Mazurek, M.~K\l{}usek-Gawenda, and A.~Szczurek,
\newblock Eur. Phys. J. A {\bf 58}, 245 (2022), 2107.13239.

\bibitem{Wang:2022gkd}
R.-j. Wang, S.~Lin, S.~Pu, Y.-f. Zhang, and Q.~Wang,
\newblock Phys. Rev. D {\bf 106}, 034025 (2022), 2204.02761.

\bibitem{Harland-Lang:2019zur}
L.~Harland-Lang, J.~Jaeckel, and M.~Spannowsky,
\newblock Phys. Lett. {\bf B793}, 281 (2019), 1902.04878.

\bibitem{Woods:1954zz}
R.~D. Woods and D.~S. Saxon,
\newblock Phys. Rev. {\bf 95}, 577 (1954).

\bibitem{Tarbert:2013jze}
C.~M. Tarbert {\em et~al.},
\newblock Phys. Rev. Lett. {\bf 112}, 242502 (2014), 1311.0168.

\bibitem{Harland-Lang:2020veo}
L.~Harland-Lang, M.~Tasevsky, V.~Khoze, and M.~Ryskin,
\newblock Eur. Phys. J. C {\bf 80}, 925 (2020), 2007.12704.

\bibitem{Bernauer:2013tpr}
A1, J.~C. Bernauer {\em et~al.},
\newblock Phys. Rev. {\bf C90}, 015206 (2014), 1307.6227.

\bibitem{Budnev:1974de}
V.~M. Budnev, I.~F. Ginzburg, G.~V. Meledin, and V.~G. Serbo,
\newblock Phys.Rept. {\bf 15}, 181 (1975).

\bibitem{Harland-Lang:2019eai}
L.~Harland-Lang,
\newblock JHEP {\bf 03}, 128 (2020), 1910.10178.

\bibitem{Harland-Lang:2010ajr}
L.~A. Harland-Lang, V.~A. Khoze, M.~G. Ryskin, and W.~J. Stirling,
\newblock Eur. Phys. J. C {\bf 69}, 179 (2010), 1005.0695.

\bibitem{Hencken:1995me}
K.~Hencken, D.~Trautmann, and G.~Baur,
\newblock Z. Phys. C {\bf 68}, 473 (1995), nucl-th/9503004.

\bibitem{Baltz:1996as}
A.~J. Baltz, M.~J. Rhoades-Brown, and J.~Weneser,
\newblock Phys. Rev. E {\bf 54}, 4233 (1996).

\bibitem{Veyssiere:1970ztg}
A.~Veyssiere, H.~Beil, R.~Bergere, P.~Carlos, and A.~Lepretre,
\newblock Nucl. Phys. A {\bf 159}, 561 (1970).

\bibitem{Lepretre:1981tf}
A.~Lepretre {\em et~al.},
\newblock Nucl. Phys. A {\bf 367}, 237 (1981).

\bibitem{Carlos:1984lvc}
P.~Carlos {\em et~al.},
\newblock Nucl. Phys. A {\bf 431}, 573 (1984).

\bibitem{Muccifora:1998ct}
V.~Muccifora {\em et~al.},
\newblock Phys. Rev. C {\bf 60}, 064616 (1999), nucl-ex/9810015.

\bibitem{Armstrong:1971ns}
T.~A. Armstrong {\em et~al.},
\newblock Phys. Rev. D {\bf 5}, 1640 (1972).

\bibitem{Armstrong:1972sa}
T.~A. Armstrong {\em et~al.},
\newblock Nucl. Phys. B {\bf 41}, 445 (1972).

\bibitem{Caldwell:1973bu}
D.~O. Caldwell {\em et~al.},
\newblock Phys. Rev. D {\bf 7}, 1362 (1973).

\bibitem{Michalowski:1977eg}
S.~Michalowski {\em et~al.},
\newblock Phys. Rev. Lett. {\bf 39}, 737 (1977).

\bibitem{Caldwell:1978ik}
D.~O. Caldwell {\em et~al.},
\newblock Phys. Rev. Lett. {\bf 42}, 553 (1979).

\bibitem{Donnachie:1992ny}
A.~Donnachie and P.~V. Landshoff,
\newblock Phys.Lett. {\bf B296}, 227 (1992), hep-ph/9209205.

\bibitem{ZEUS:2001wan}
ZEUS, S.~Chekanov {\em et~al.},
\newblock Nucl. Phys. B {\bf 627}, 3 (2002), hep-ex/0202034.

\bibitem{Baur:1986uso}
G.~Baur and C.~A. Bertulani,
\newblock Phys. Lett. B {\bf 174}, 23 (1986).

\bibitem{Sakurai:1969ss}
J.~J. Sakurai,
\newblock Phys. Rev. Lett. {\bf 22}, 981 (1969).

\bibitem{Sakurai:1960ju}
J.~J. Sakurai,
\newblock Annals Phys. {\bf 11}, 1 (1960).

\bibitem{Bauer:1977iq}
T.~H. Bauer, R.~D. Spital, D.~R. Yennie, and F.~M. Pipkin,
\newblock Rev. Mod. Phys. {\bf 50}, 261 (1978),
\newblock [Erratum: Rev.Mod.Phys. 51, 407 (1979)].

\bibitem{Li:2019sin}
C.~Li, J.~Zhou, and Y.-J. Zhou,
\newblock Phys. Rev. D {\bf 101}, 034015 (2020), 1911.00237.

\bibitem{Shao:2022stc}
D.~Y. Shao, C.~Zhang, J.~Zhou, and Y.~Zhou,
\newblock (2022), 2212.05775.

\bibitem{Xiao:2020ddm}
B.-W. Xiao, F.~Yuan, and J.~Zhou,
\newblock Phys. Rev. Lett. {\bf 125}, 232301 (2020), 2003.06352.

\bibitem{Khoze:2002dc}
V.~A. Khoze, A.~D. Martin, and M.~G. Ryskin,
\newblock Eur.Phys.J. {\bf C24}, 459 (2002), hep-ph/0201301.

\bibitem{Zha:2018tlq}
W.~Zha, J.~D. Brandenburg, Z.~Tang, and Z.~Xu,
\newblock Phys. Lett. B {\bf 800}, 135089 (2020), 1812.02820.

\bibitem{Brandenburg:2021lnj}
J.~D. Brandenburg, W.~Zha, and Z.~Xu,
\newblock Eur. Phys. J. A {\bf 57}, 299 (2021), 2103.16623.

\end{thebibliography}
\bibliographystyle{h-physrev}

\end{document}